\documentclass[a4paper,11pt]{article}
\usepackage{jinstpub} 
\usepackage{lineno}

\usepackage{siunitx}
\usepackage{caption}
\usepackage{subcaption}
\usepackage{longtable}

\title{EASpy: Fast simulation of fluorescence and Cherenkov light from extended air showers at large zenith angles}

\author[1]{Ali Baktash \note{Corresponding author.}}
\author{and Dieter Horns}
\affiliation{University of Hamburg,\\
Faculty of Mathematics, Informatics and Natural Sciences,\\
Institut fuer Experimentalphysik,\\
Luruper Chaussee 149, 22761 Hamburg}

\emailAdd{ali.baktash@uni-hamburg.de}
\abstract{
The detailed simulation of extended air showers (EAS) 
and their emission of Cherenkov and fluorescence light 
requires increasing computation time and storage volume with increasing energy of the primary
particle. Given these limitations, it is currently challenging to optimize configurations of imaging air Cherenkov telescopes at photon energies beyond approximately 100 TeV. Additionally, the existing simulation frameworks are not capable of capturing the interplay of Cherenkov and fluorescence light emission at large zenith
angle distances ($\gtrsim 70^\circ$), where the collection area of Cherenkov telescopes considerably increases.
Here, we present  \texttt{EASpy}, a framework for the simulation of EAS at large zenith angles
using parametrizations for electron-positron distributions.
Our proposed approach for the emission of fluorescence and Cherenkov light and the subsequent imaging of these components by Imaging Atmospheric Cherenkov Telescopes (IACTs) aims to provide flexibility and accuracy while at the same time it reduces the computation time considerably compared to full Monte Carlo simulations. 
We find excellent agreement of the resulting Cherenkov images when comparing results
obtained from \texttt{EASpy} with the de-facto standard simulation tool \texttt{CORSIKA} and
\texttt{sim\_telarray}. 
In the process of verifying our approach, we have found that air shower images appear wider and longer with increasing impact distance at large zenith angles, an effect that has previously not been noted. We also investigate the distribution of light on the ground for fluorescence and Cherenkov emission and highlight their key differences to distributions at moderate zenith angles.
}


\arxivnumber{TBD} 

\begin{document}
\maketitle
\flushbottom

\section{Introduction}
\label{sec:intro}

Extended air showers (EAS) provide a unique opportunity to study cosmic rays with ground based observations. At increasing energy, the sensitivity of particle detectors, e.g. Imaging Atmospheric Cherenkov Telescopes (IACTs), to observe the produced Cherenkov and fluorescence light from an EAS depends primarily on the collection area $A_{\mathrm{eff}}$. 
As a consequence of the limited collection area of the current generation of IACTS, the gamma-ray sky at energies  above \SI{100}{\tera\electronvolt} remains largely unexplored. 
The photon rate from the Crab Nebula, one of the brightest steady Galactic gamma-ray sources,  
for a  collection area of $A_\mathrm{eff}=\SI{1}{\square\kilo\metre}$ is less than one photon 
per \SI{100}{\hour} \cite{2023A&A...671A..67D}. One way to increase $A_{\mathrm{eff}}$ is to use multiple IACTs ($\sim 100$) commonly placed with a distance of the order of $\sim$ \SI{100}{m} apart from each other \cite{2011ExA....32..193A}. A more budget-friendly approach is to make use of fewer stand-alone telescopes observing at large zenith angles (LZA), which typically means zenith angles larger than \SI{70}{\degree}. With observations at LZA the distance to the shower is greatly increased (\SIrange[]{50}{100}{\kilo\metre}) which consequently expands the light pool on the ground \cite{Sommer_et_al} (albeit at the cost of increased attenuation due to a larger photon path to the observer). The MAGIC collaboration has already successfully demonstrated
that observations at a zenith angle range of \SIrange{70}{80}{\degree} increase the collection area to be larger than a square km. With  an observation time of $\sim$ \SI{56}{\hour},
they were able to detect $\gamma$-ray emission from the Crab Nebula up to energies of 100~TeV \cite{MAGIC_LZA_CRAB2020}.

So far, the operational and planned IACT arrays have been optimized for observations at small
zenith angles. The optimization of their configuration at large zenith angles requires a substantial effort in simulating a large number of air showers at high energies and their detection with a variety of telescope layouts, camera sizes, mirror diameters, etc. 
In the context of performance studies, Monte Carlo simulations, e.g. \texttt{CORSIKA} (see Appendix~\ref{subsec:corsika}), have become essential in order to optimize IACTs and reconstruction methods. Although \texttt{CORSIKA} is not able to simulate fluorescence light, there has been work done by the community to implement fluorescence emission 
in \texttt{CORSIKA} (\cite{DESOUZA2004263}, \cite{Morcuende_2019}). A general downside of these Monte Carlo simulations is the increased computation time with higher primary particle energies (and to an lesser extend increased storage size).

A simplified approach to make simulations more time-efficient is to use parametrizations which describe the distribution of charged particles in an EAS rather than simulating the shower development and keeping track of each particle generated. Commonly used analytical functions for the longitudinal number of charged particles in an EAS are the Greisen \cite{Greisen} or Gaisser-Hillas \cite{Gaisser_Hillas} function and for the lateral distribution the Nishimura-Kamata-Greisen function \cite{Nishimura_Kamata}. The former mentioned functions together with parametrizations of fluorescence and Cherenkov light emission \cite{Nerling_2006} have been used in \texttt{ShowerModel} \cite{jaime_rosado_2022_6773258, Morcuende:2021bmo} to simulate the light emission in EAS and its detection by IACTs. 
While parametrizations tend to miss the fluctuations in the shower development, this is less of a problem at higher primary particle energies. Here, the fluctuations are considerably smaller than at lower energies. 

In this work we present \texttt{EASpy} \footnote{https://github.com/4liBaktash/EASpy}, a 3.5-dimensional simulation of EAS based on parametrizations for electron-positron distributions and its consequent emission of fluorescence and Cherenkov light in a analytical way while taking into account the curvature of Earth's atmosphere. As in \texttt{ShowerModel}, our simulation aims to compute the detector response in a fast pace but at the same time reach a high level of accuracy and flexibility. Here, we focus on 
photon-initiated showers to establish the method and compare the results obtained using
\texttt{EASpy} with \texttt{CORSIKA} and \texttt{sim\_telarray}, independent of
particular hadronic interaction models. 

One of the reasons why \texttt{ShowerModel} is not applicable for LZA observations is that it does not take into account the curvature of Earth's atmosphere, which can not be neglected anymore for the zenith angle range discussed in this work. In Table~\ref{tab:intro_comparison}, we provide an overview of some properties for \texttt{CORSIKA}/\texttt{sim\_telarray} (see Appendix~\ref{subsec:corsika}/\ref{subsec:simtel}), \texttt{ShowerModel}, and \texttt{EASpy}. While the simplified approach for the light collection is independent of the pixel shape, \texttt{EASpy} does take into account  gaps between the pixels in case of round pixels.
Furthermore, in \cite{DESOUZA2004263} the authors showed how for core distances lower than \SI{8}{\kilo\metre} the detector response of one-dimensional and three-dimensional simulations is quite different in the lateral (transversal) spread of the resulting fluorescence light image. An additional motivation for this work was to not use any parametrization for Cherenkov and fluorescence light emission, since in principle the distribution of charged particles is sufficient to treat both emission types in a analytical way. 

\begin{table}[h!]
    \caption{Comparison of properties between \texttt{CORSIKA}/\texttt{sim\_telarray}, \texttt{ShowerModel}, \texttt{EASpy}.}
    \label{tab:intro_comparison}
    \begin{center}
\begin{tabular}{|l|ccc|}
\hline
  & \texttt{CORSIKA}/\texttt{sim\_telarray} & \texttt{ShowerModel} & \texttt{EASpy} \\
 \hline
 Simulation method & Full MC & parameterized & parameterized \\
 Dimension (space+time) & 3+1 & 1+1 & 2.5+1 \\
Atmosphere & (sliding) plane parallel & plane parallel  & spherical  \\
Particle content & $e^{\pm}$, $\mu^{\pm}$, $\nu$, hadrons & $e^{\pm}$ & $e^{\pm}$ \\ 
Fluorescence & no & yes (param.) & yes (particle based) \\
Cherenkov & yes (particle based) & yes (param.) & yes (particle based) \\
Cherenkov-light collection & ray-tracing & parametrized & geometrical approach \\
Zenith angle & \SI{0}{\degree} -- \SI{90}{\degree} & \SI{0}{\degree} -- \SI{70}{\degree} & \SI{70}{\degree} -- \SI{90}{\degree} \\
Imaging: PSF & yes & no & yes \\
Imaging: Pixel shape & hex./square/circ. & square & hex./square/circ. \\
\hline
\end{tabular}
\end{center}
\end{table}

In Section~\ref{sec:framework}, we introduce the basic concepts of \texttt{EASpy} including the production of Cherenkov and fluorescence light from an EAS and the consequent collection of this light by an IACT. The simulation of the detector response is explained in Section~\ref{sec:imaging} together with a comparison of shower images obtained with \texttt{EASpy} and a full air-shower simulation in Section~\ref{sec:verification_with_full_simulation}. In Section~\ref{sec:ground_distribution}, we finally discuss the characteristics of the Cherenkov and fluorescence photon ground distribution at LZA.

\section{EASpy framework}
\label{sec:framework}

\subsection{Spherical Atmosphere}
\begin{figure}[htbp!]
     \centering
     \begin{subfigure}[c]{0.49\textwidth}
     \centering
     \includegraphics[]{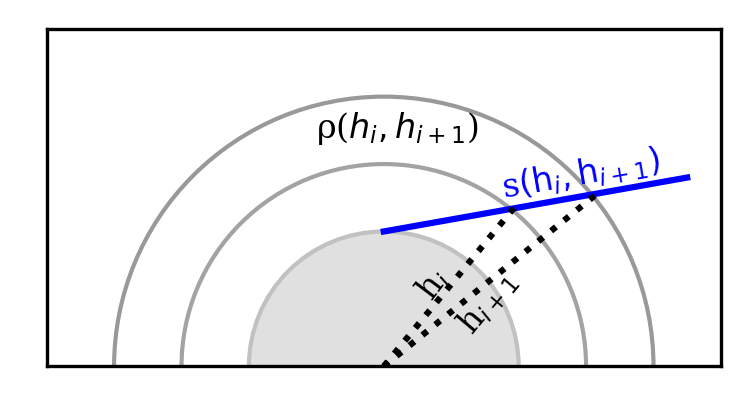}
     \caption{}
     \label{fig:spherical_atmo}
     \end{subfigure}
     \hfill
     \begin{subfigure}[c]{0.49\textwidth}
      \centering
      \includegraphics[]{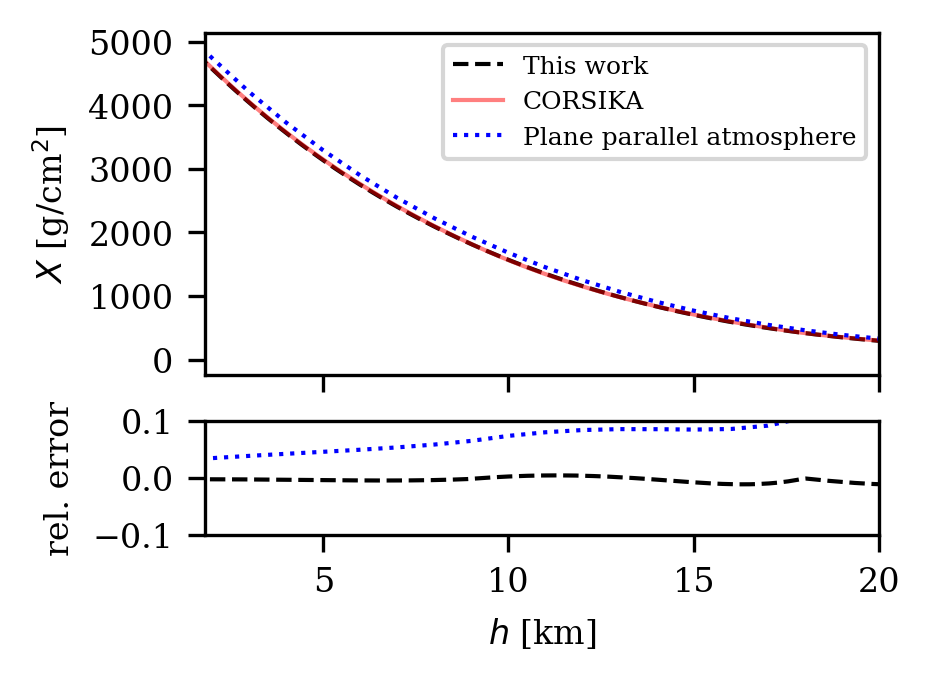}
      \caption{}
     \label{fig:slant_depth}
     \end{subfigure}
\caption{(a) Model of spherical atmosphere. Gray shaded area denotes the Earth, the blue line marks the path of the primary particle and the dashed lines the height of the bin edges for one spherical shell measured from the midpoint of the earth. The traversed slant distance for one spherical shell is then $s(h_{i}, h_{i+1})$.
(b) Upper panel: Slant depth as a function of height at a zenith angle of 80\textdegree. Lower panel: Relative error for plane parallel atmosphere and this work compared to \texttt{CORSIKA}.}
\end{figure}

Since the fluorescence and Cherenkov light yields depend on atmospheric parameters (e.g., air density, temperature, pressure, ...) it is important to choose a suitable model for the atmosphere. Typically, at zenith angles $\leq$ \SI{70}{\degree} a plane-parallel atmosphere is assumed, i.e. that the air mass scales with the secant of the zenith angle. However, at zenith angles $\geq$ \SI{70}{\degree} the curvature of the atmosphere is not negligible anymore and a plane parallel atmosphere approximation will lead to large errors for, e.g., the slant depth and atmospheric transmission.
In order to have a more precise model, the atmosphere is divided into a series of equally spaced spherical shells with 
a separation of \SI{6}{\metre}. The slant distance travelled and the height of each spherical shell is calculated in a coordinate system where the origin is set to the midpoint of the Earth (see Fig. \ref{fig:spherical_atmo}). 

With measurements for the air density at various heights at zenith = \SI{0}{\degree} one can use the spherical symmetry of the model and calculate the height dependent slant depth at LZA:
\begin{equation}
\label{eqn:slant_depth}
    X(h_{i}, h_{i+1}) = \int\limits_{h_{i}}^{h_{i+1}} \rho(h)~\mathrm{d}s = \overline{\rho(h_{i}, h_{i+1})}\,s(h_{i}, h_{i+1}),
\end{equation}
where $X(h_{i}, h_{i+1})$ is the traversed slant depth for one spherical shell with bin edges at heights $h_{i}$ and $h_{i+1}$, $\overline{\rho(h_{i}, h_{i+1})}$ is the 
(linearly) averaged air density and $s(h_{i}, h_{i+1})$ is the slant distance for the primary particle path between the bin edges\footnote{with $\rho\propto \exp(-h/h_0)$ and $h_0\gg \Delta h$, the
linear approximation for $\rho$ in a shell is sufficiently accurate}. This way, the atmospheric parameters along the path of the primary particle can be calculated.  In this approximation, we can treat the shower to be symmetric in azimuth. In Fig.~\ref{fig:slant_depth}, a comparison between \texttt{CORSIKA} (compiled with the "CURVED" option), plane parallel atmosphere and this work is shown for the height dependent slant depth at a zenith angle of \SI{80}{\degree}. One can observe that the results from \texttt{CORSIKA} and this work are in good agreement while for a plane parallel atmosphere model the relative error compared to the \texttt{CORSIKA} results is $\simeq$ 5\% at observation level and increasing with height.\\

\subsection{Atmospheric transmission}
The atmospheric transmission for Cherenkov and fluorescence light generated by air showers
at large zenith angles is smaller than at small zenith angles. Since the position of the shower maximum is observed at a distance of 
\SIrange{50}{100}{\kilo\metre},  more light will be scattered and absorbed in the atmosphere.
The resulting transmission has been calculated using the \texttt{MODTRAN} \cite{MODTRAN} 
program which takes into account  Rayleigh- and Mie-scattering as well as absorption processes
with a moderate spectral resolution. The program offers a wide choice of pre-defined 
atmospheric conditions. We use the tropical atmosphere with a desert-type haze (aerosol) 
condition. 

\begin{figure}[htp!]
\centering
\includegraphics[]{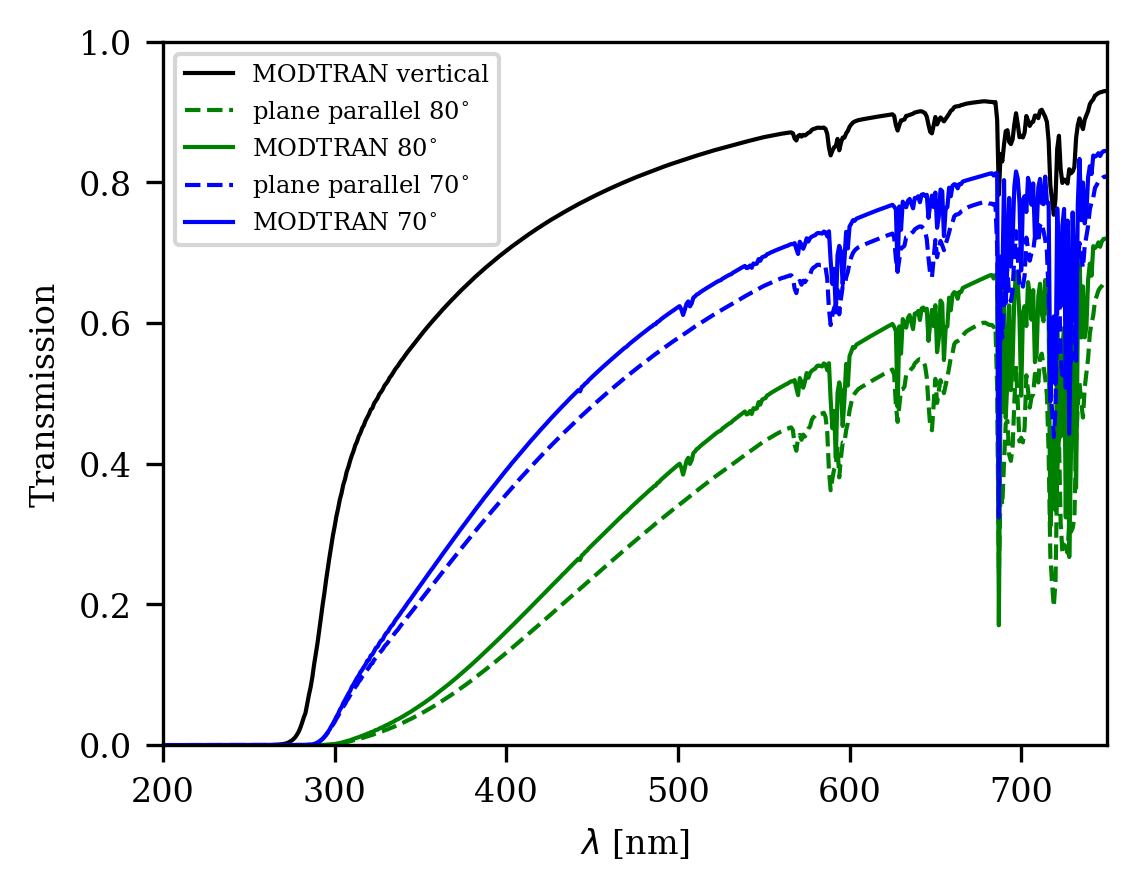}
\caption{Transmission from \SI{15}{\kilo\metre} to \SI{1.8}{\kilo\metre} height for a range of wavelengths $\lambda$ for a vertical path, \SI{70}{\degree} and \SI{80}{\degree} zenith angle. Transmission profiles calculated with \texttt{MODTRAN} are for tropical atmospheric profile and desert haze.}
\label{fig:transmission}
\end{figure}

In a plane parallel atmosphere model one would calculate the transmission $T$ for a vertical path and assume that $\tau \propto \sec(\theta)$, where $\tau$ is the optical depth with $T = \exp(-\tau)$. In comparison to a spherical atmosphere, this approach would overestimate the
mass overburden resulting in an underestimate of the transmission.
In Fig. \ref{fig:transmission},
one can observe that a plane parallel atmosphere model would underestimate the transmission by $\sim 10\%$ and that the transmission for a wavelength of $\sim$ \SI{500}{\nano\metre} is reduced by a factor of 2 for a path with \SI{80}{\degree} zenith angle compared to a vertical path. Furthermore, as expected, the difference between a plane parallel atmosphere approximation and the \texttt{MODTRAN} results is lower for a path with \SI{70}{\degree} zenith angle compared to \SI{80}{\degree} zenith angle.

The simulation tool provides a choice between using the plane-parallel approximation or 
to use a  table of the transmission for different values of zenith angle. The EASpy package will include a few selected look-up tables for the transmission.

\subsection{Parametrizing the
electromagnetic shower component}
\label{sec:parametrizing}
In order to simulate fluorescence and Cherenkov light from air showers one needs to accurately describe the distribution of charged particles, and most importantly the distribution of electrons and positrons since these particles are the most abundant charged particles in air showers. In this work we use a multi-dimensional parametrization for the electron-positron distributions for particle energy, angular spectrum and lateral distance proposed by \cite{LAFEBRE2009243}. 

The Ansatz for the parametrization
of the differential 
particle number $n$ in a logarithmic interval [$\ln\epsilon, \ln\epsilon+\mathrm{d}\ln\epsilon]$ 
of kinetic energy $\epsilon$ of electrons and positrons in MeV follow from Eqn.~6  in Ref.~\cite{LAFEBRE2009243}
\begin{equation}
\label{eq:para_energy}
n(t) \coloneqq
\frac{1}{N(t)} \frac{\partial N(t)}{\partial \ln\epsilon} =
\frac{1}{N(t)} N(t; \ln\epsilon),
\end{equation}
for the particle number $n_\Omega$ in a differential interval of 
solid angle $[\Omega,\Omega + \mathrm{d}\Omega]$ given in Eqn.~8 in Ref.~\cite{LAFEBRE2009243}
\begin{equation}
\label{eq:para_angular}
n_\Omega(t) \coloneqq \frac{1}{N(t; \ln\epsilon)} \frac{\partial^2N(t)}{\partial\ln\epsilon \partial\Omega},
\end{equation}
and for the particle number $n_x$ in a differential logarithmic interval
of radial distance to the shower axis [$\ln x$, $\ln x + \mathrm{d}\ln x$], where  $x \equiv \frac{r}{r_{\mathrm{M}}}$ and $r_{\mathrm{M}}$ is the Moli\`{e}re radius
\begin{equation}
\label{eq:para_lateral}
n_x(t) \coloneqq \frac{1}{N(t; \ln\epsilon)} \frac{\partial^2N(t)}{\partial\ln\epsilon \partial\ln x}.
\end{equation}

The distributions are fully described in terms of relative evolution stage t
\begin{equation}
\label{eq:relative_evolution_stage_t}
    t = \frac{X - X_{\mathrm{max}}}{X_{0}},
\end{equation}
where the maximum number of particles is reached at slant depth $X_{\mathrm{max}}$ and $X_{0} \simeq 36.7~\mathrm{g~cm^{-2}}$ is the radiation length of electrons and positrons in air.

Here, $n(t)$, $n_{\Omega}(t)$ and $n_{x}(t)$ are probability density functions given by Eqn.~6, 8 and 14 in Ref.~\cite{LAFEBRE2009243}. In order to get the number of particles in the range of [$\ln\epsilon, \ln\epsilon+\mathrm{d}\ln\epsilon$] and [$\ln x, \ln x+\mathrm{d}\ln x$] at relative evolution stage $t$ we calculate:
\begin{equation}
\label{eq:true_particle_number}
    N(t, \bar\epsilon, \bar x) = N(t) \iint
    n_{x}(t) ~\mathrm{d}\ln\epsilon~\mathrm{d}\ln x,
\end{equation}
where $N(t)$ is the total number of electrons-positrons crossing a plane at level $t$ perpendicular to the shower axis. The values for $\bar x$ and $\bar \epsilon$ are chosen to be the arithmetic and geometric averages of the intervals [$\ln\epsilon, \ln\epsilon+\mathrm{d}\ln\epsilon$] and [$\ln x, \ln x+\mathrm{d}\ln x$] respectively. Note, to keep the 
notation simple, we will use in the following $\epsilon$ and $x$ instead of $\bar \epsilon$
and $\bar x$.

\subsection{Electron-positron distributions}\label{subsec:electron-positron_distribution}
\begin{figure}[htp!]
\begin{center}
\includegraphics[width=0.46\linewidth]{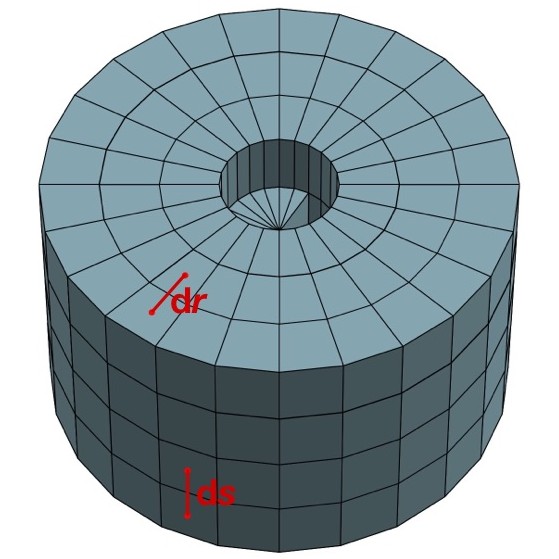}
\caption{Schematic sketch of shower binning.}
\label{fig:shower_voxels}
\end{center}
\end{figure}

Given a shower profile $N(t)$ which describes the total number of particles at shower evolution stage $t$, we use Eqn.~\ref{eq:true_particle_number} to determine
the number of particles in the differential intervals for distance $r$ and energy $\epsilon$: $N(t, \epsilon, r/r_{M})$.  Therefore the air shower is assumed to have a cylindrical symmetry and is binned in the following way: For a fixed energy range of [$\ln\epsilon, \ln\epsilon+\mathrm{d}\ln\epsilon]$ the air shower is binned along the shower axis with bin width d$s$ (see Eqn.~\ref{eqn:slant_depth}) perpendicular to the shower axis with radial bin width d$r$. The resultant cylindrical shells are then subdivided into $N_{\phi}$ equally spaced bins (so-called "voxels") over $2\pi$ around the shower axis (see Fig.~\ref{fig:shower_voxels}). For each of these voxels, the energy-dependent number of particles is given by $N(t, \epsilon, r/r_{M})/N_{\phi}$.

\begin{figure}[htbp!]
     \centering
     \begin{subfigure}[c]{0.49\textwidth}
     \centering
     \includegraphics[]{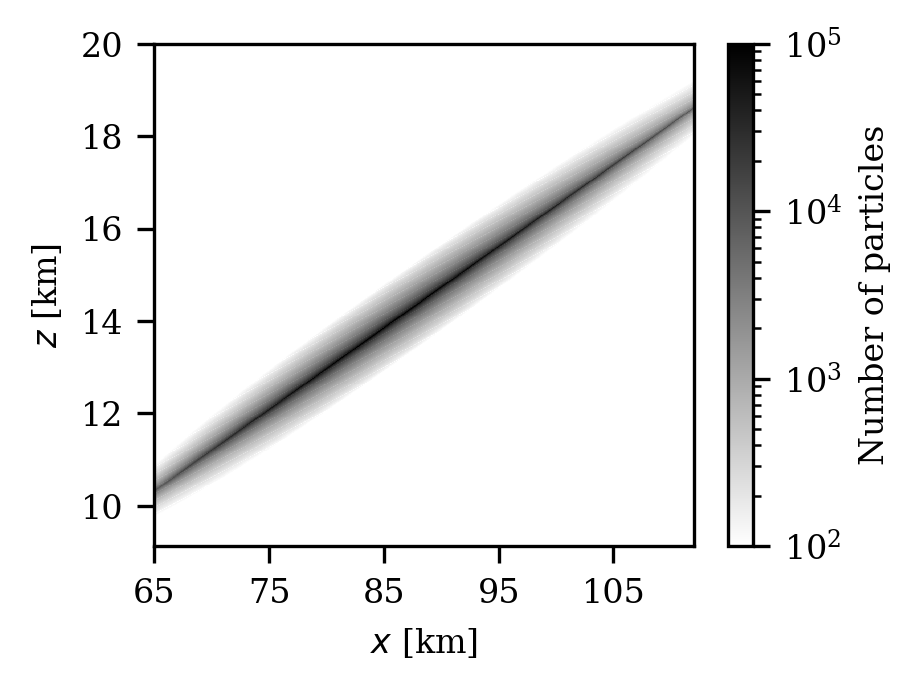}
     \caption{}
     \label{fig:particle_distribution}
     \end{subfigure}
     \hfill
     \begin{subfigure}[c]{0.49\textwidth}
      \centering
      \includegraphics[]{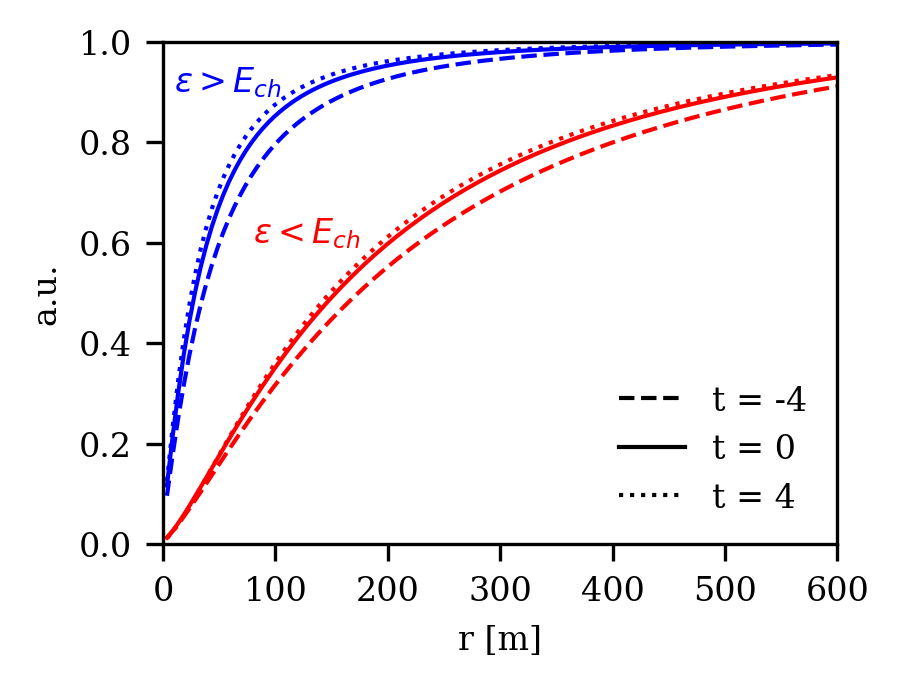}
      \caption{}
     \label{fig:energy_dependent_dist}
     \end{subfigure}
\caption{(a) Electron and positron distribution projected on the $xz$ plane (i.e., integrated over y-coordinate) for a photon initiated air shower with primary energy of $E_{\gamma}$ = \SI{975}{\tera\electronvolt} at a zenith angle of \SI{80}{\degree}.
(b) Normalized cumulative distribution of particle numbers as a function of distance to the shower axis at $t = -4, 0, 4$ for the same air shower as in \ref{fig:particle_distribution}. $E_{\mathrm{Ch}}$ denotes the Cherenkov energy threshold. The maximum distance to the shower axis is set to \SI{1}{\kilo\metre}.}
\end{figure}

For an overview of the shower development, we show in
 Fig. \ref{fig:particle_distribution} the distribution of electrons and positrons (integrated over the energy range of $\epsilon_\mathrm{min}=$\SI{1}{\mega\electronvolt} to
 $\epsilon_\mathrm{max}=$\SI{10}{\giga\electronvolt}) projected on the $xz$ plane for a photon initiated air shower with zenith angle of \SI{80}{\degree}. One can clearly observe the shower axis together with the shower maximum at around  $x \simeq$ \SI{85}{\kilo\metre} and $z \simeq$ \SI{14}{\kilo\metre}. 
 
The lateral distribution of shower particles around the shower axis will contribute 
to the production of Cherenkov light if their energy exceeds the energy threshold of 
$E_{Ch} \approx \SI{50}{\mega\electronvolt}$ at the position close to the shower maximum.
In Fig.~\ref{fig:energy_dependent_dist}, we compare the lateral distribution for particles
emitting Cherenkov light ($\epsilon>E_{Ch}$)
and those which will produce  fluorescent light only ($\epsilon<E_{Ch}$). A large fraction
($80~\%$) of the
Cherenkov-light emitting particles are confined within \SI{75}{\metre} while
fluorescent-light emitting particles are distributed over a larger distance: 80~\% 
of these particles are located within \SI{400}{\metre} to the shower axis. 
 
\subsection{Ionization energy deposit}\label{subsec:Edep}
Electrons and positrons deposit energy differently in collisions with air molecules when passing through the atmosphere \cite{SELTZER19821189}. The stopping power for electrons can be calculated with the M\o ller cross section and for positrons with the Bhabha cross section \cite{doi:10.1146/annurev.ns.04.120154.001531}. Both stopping power formulas including the parameter values used in this work can be found in \cite{PDG}. The resulting ionization energy losses are assumed to be deposited at the midpoints of each voxel. Note, that in our framework the assumed cylindrical symmetry of the particle distribution defines the symmetry of the energy deposit. This is an approximation, since at LZA the atmospheric conditions around the shower axis for fixed $t$ and $r$ will deviate slightly for different azimuth angles. This approximation is however sufficient to capture the main features of the air showers relevant for the resulting image. This is demonstrated when comparing EASpy generated images with the full simulation as shown in Section~\ref{sec:verification_with_full_simulation}.

We can readily estimate the accuracy of
 the parametrization for the energy spectrum given in Eqn.~\ref{eq:para_energy} by 
 calculating the energy deposit using the stopping powers for electrons and positrons and compare it with results obtained with \texttt{CORSIKA} (see Appendix~\ref{subsec:corsika}).
 In Fig.~\ref{fig:dEdX}, we compare our results for the energy deposit per slant depth $\frac{\mathrm{d}E}{\mathrm{d}X}$ with  \texttt{CORSIKA} for photon initiated air showers in the energy range of \SIrange{900}{1000}{\tera\electronvolt} at a zenith angle of \SI{70}{\degree} and \SI{80}{\degree}. The number of individual air showers is $\sim$ 850 per zenith angle. The relative error is calculated as
 
\begin{equation*}
    \mathrm{rel.\,error} = \frac{(\mathrm{d}E/\mathrm{d}X)_{\texttt{EASpy}} - (\mathrm{d}E/\mathrm{d}X)_{\mathrm{\texttt{CORSIKA}}}}{(\mathrm{d}E/\mathrm{d}X)_{\mathrm{\texttt{CORSIKA}}}}.
\end{equation*}

\begin{figure}[htp!]
\centering
\includegraphics[width=0.7\linewidth]{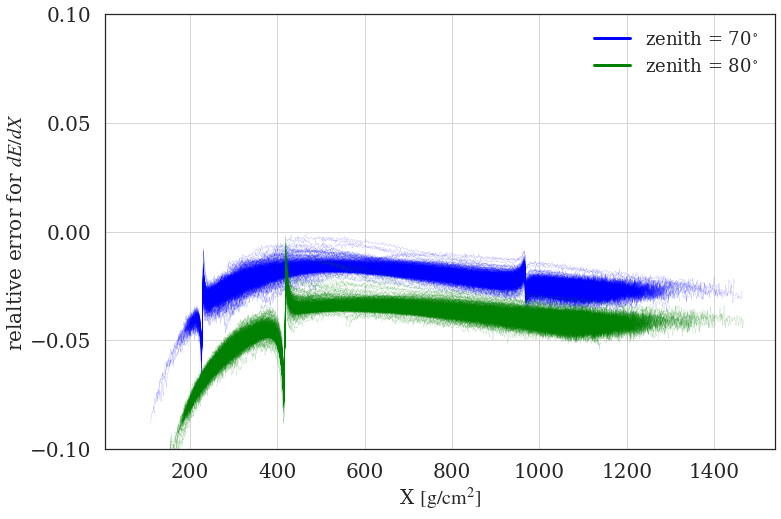}
\caption{Relative error for the energy deposit per slant depth $\frac{\mathrm{d}E}{\mathrm{d}X}$ as a function of slanth depth $X$ for photon initiated air showers with zenith angles \SI{70}{\degree} and \SI{80}{\degree}.}
\label{fig:dEdX}
\end{figure}

\noindent One can observe that at early shower stages the relative difference is $\sim 5-10\%$ while at later stages the correspondence is better than $5\%$ for both zenith angles. We found that deviations from the \texttt{CORSIKA} results mainly depend on the zenith angle, i.e., a slightly better correspondence at zenith angles around $\sim$ \SI{70}{\degree} compared to zenith angles closer to \SI{80}{\degree} and a negligible dependence on the primary particle energy. 

The "spikes" in Fig. \ref{fig:dEdX} are related to the way the atmosphere is treated with
 \texttt{CORSIKA}: The "CURVED" option uses a "sliding plane atmosphere" instead of a fully spherical system. Every time the horizontal displacement of a particle exceeds a limit of \SIrange{6}{20}{\kilo\metre} (depending on altitude), a transition to a new, locally plane atmosphere is performed \cite{CORSIKA_users_guide}. Apparently this treatment introduces "spikes" whenever the transition is performed. 

\subsection{Fluorescence light}
A fraction of the ionization energy deposited in
the atmosphere (see \ref{subsec:Edep}) leads to the
excitation of the nitrogen atoms with subsequent
fluorescence light emission with a typical life time of the excited state of order $\sim$ \SI{40}{\nano\second} (e.g. \cite{Arqueros_2009}, \cite{Itikawa_et_al}). 
The number of produced fluorescence photons $N^{fl}$ 
in a voxel at position $(t, r, \phi)$ can be calculated with:

\begin{equation}\label{eq:dNdX_fl}
    N^{fl}(t, r, \phi) = \frac{\Delta X(t)}{N_\phi}                                  \int\limits_{\epsilon_{\mathrm{min}}}^{\epsilon_{\mathrm{max}}} 
                             N(t,\epsilon,r/r_M)  
                             \frac{\mathrm{d}E}{\mathrm{d}X}(\epsilon)
                             \mathrm{d}\epsilon \cdot
                             \int\limits_{\lambda_{\mathrm{min}}}^{\lambda_{\mathrm{max}}}               
                             Y_{air}(\lambda, T, p, p_{w}) \mathrm{d}\lambda,
\end{equation}

\noindent where $\Delta X(t)$ is the traversed slant depth depending on relative evolution stage $t$ and $Y_{air}(\lambda, T, p, p_{w})$ is the fluorescence yield in air depending on wavelength $\lambda$ (in nm),
 ambient atmospheric 
pressure $p=p(t)$, water vapour partial pressure $p_{w}=p_w(t)$ and temperature $T=T(t)$. 
This way, the amount of fluorescence light is
directly proportional to the rate of ionization energy loss $\frac{\mathrm{d}E}{\mathrm{d}X}$.
 The fluorescence yield in air is usually 
 expressed in terms of the absolute yield of the \SI{337}{\nano\metre} band in dry air at
 reference pressure $p_{0}$ and temperature $T_{0}$:
 
\begin{align*}
     \frac{Y_{air}(\lambda, T, p, p_{w})}{Y_{air}(337, T_{0}, p_{0})} &=  
     \frac{I_{\lambda}(T_{0}, p_{0})}{I_{337}(T_{0},p_{0})} 
                            \cdot \frac{1+ p_{0}/p^{'}_{337}(T_{0})}{1+p/p^{'}_{\lambda}(T, p_{w})},
\end{align*}

\noindent where $I_{\lambda}(p_{0}, T_{0})$ is the 
specific  intensity and the function $p^{'}_{\lambda}(T, p_{w})$
takes into account temperature-dependent 
non-radiative de-excitation effects. Since the absolute yield $Y_{air}(337,T_0,p_0)$ is given for 
a dry atmosphere, it is assumed that the reference water vapor pressure $p_{w,0}=0$. Similarly, the values $I_\lambda(T_0, p_0)$, $I_{337}(T_0,p_0)$ and
$p'_{337}(T_0)$ are given for a dry atmosphere.

\begin{figure}[htp!]
\centering
\includegraphics[]{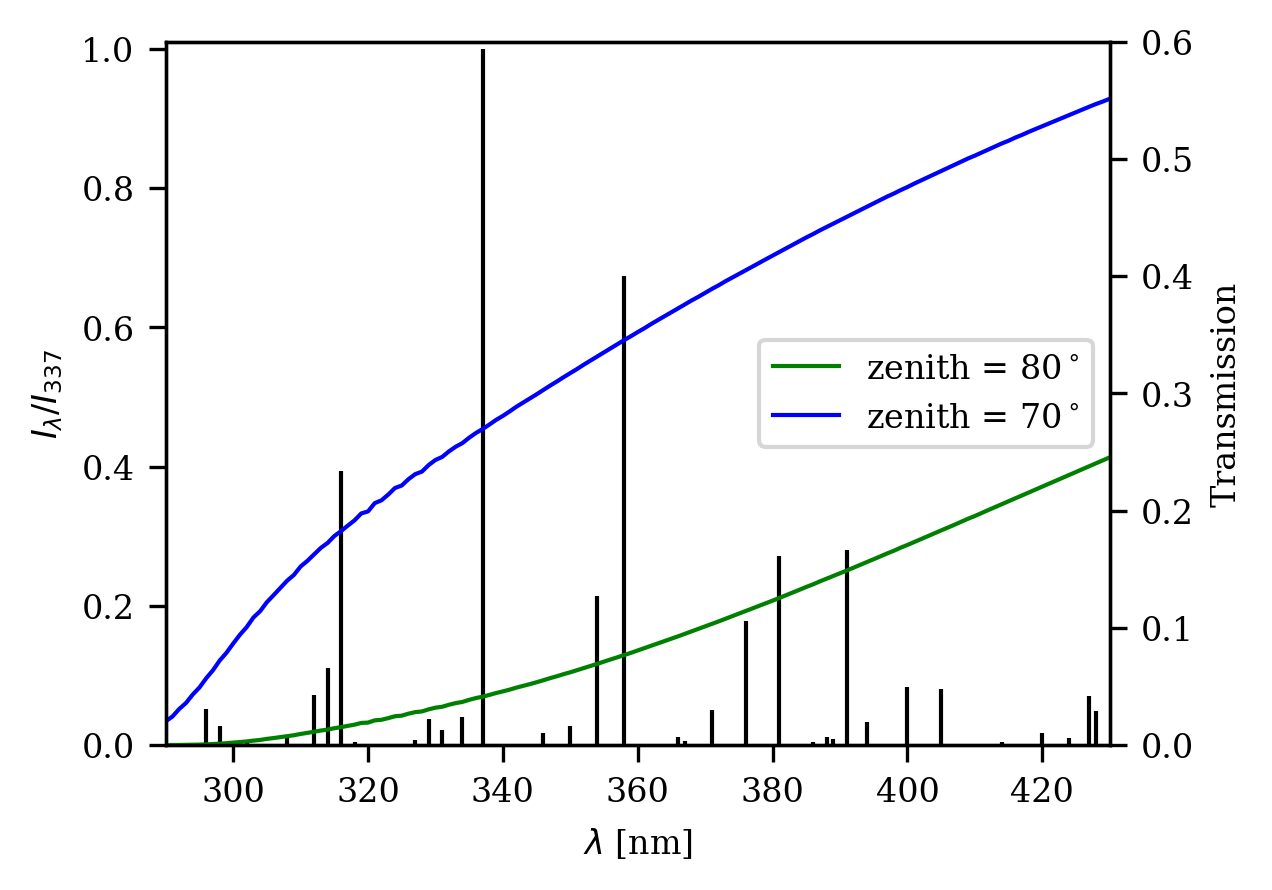}
\caption{Black lines denote the wavelength dependent intensity relative to the \SI{337}{\nano\metre} band $I_{\lambda}/I_{337}$ (left axis). The blue(\SI{70}{\degree}) and green(\SI{80}{\degree}) lines describe the transmission from the position of the shower maximum down to the observation level (right axis).}
\label{fig:rel_intensity_vs_trans}
\end{figure}

The
relative intensity $I_{\lambda}/I_{337}$ (see Fig.~\ref{fig:rel_intensity_vs_trans}) highlights that most
of the fluorescence lines are located at $\lambda\lesssim$ \SI{400}{\nano\metre}.
A detailed discussion on the temperature, humidity and pressure dependence of the air-fluorescence yield and the function  $p^{'}_{\lambda}(T, p_{w})$ can be found in \cite{Ave_2007,Ave_2008}. The parameters used 
for the air-fluorescence yield are listed in Appendix~\ref{sec:appendix_FL}. Note, that \texttt{EASpy} calculates the atmospheric parameters only along the shower axis. At relative evolution stage $t$ all voxels in a plane perpendicular to the shower axis will have the same values for the atmospheric parameters.

We assume that  fluorescence emission is isotropic. Therefore, only 
a small fraction of the 
fluorescence photons will be observed by the telescopes.
Instead of randomly distributing the photons over 
$4\pi$ and keeping track of each photon trajectory we assume 
that the  amount of fluorescence photons $N^{fl}_{sphere}$ hitting the 
telescope's sphere is given by its angular size seen from the position of a  
individual voxel (i.e., the telescope is modelled as a sphere with radius set to the radius of the mirror, see Appendix~\ref{subsec:simtel}). Consequently, the fraction $N^{fl}_{sphere}/N^{fl}$ of the 
produced 
fluorescence photons hitting the telescope's sphere   depends 
only  on the ratio of its radius $R$ and  distance  $d$ to the voxel position:
\begin{equation}
\begin{aligned}
    \frac{N_{sphere}^{fl}}{N^{fl}}&=  
    \frac{\Delta \Omega(d,R)}{4\pi} \\
                     &= \frac{ 1 - \sqrt{1 - R^2/d^2}}{2}\approx \frac{R^2}{4d^2}.
\end{aligned}
\end{equation}
For a zenith angle of \SI{70}{\degree}(\SI{80}{\degree}) and a radius $R =$ \SI{14}{\metre}, the resulting fraction is $\approx 10^{-8}(10^{-9})$. 
 This substantial decrease when going from \SI{70}{\degree} to \SI{80}{\degree} is a direct consequence of the fact that the distance of the telescope to the shower maximum increases rapidly with increasing zenith angle of the air shower. 
 
 In addition to the geometrical dilution of the fluorescence light, the atmospheric transmission decreases with increasing zenith angle (increasing slant depth). 
 We compare the relative transmissivity for zenith angles of \SI{70}{\degree} and \SI{80}{\degree} in Fig.~\ref{fig:rel_intensity_vs_trans}.
  The starting height is taken as the average height of 
  the shower maximum of air showers with $E_{\gamma} \sim$ \SI{1}{\peta\electronvolt} (\SI{70}{\degree}: $\sim$ \SI{10}{\kilo\metre} and \SI{80}{\degree}: $\sim$ \SI{14}{\kilo\metre}).
  The observation level is at a height of $\sim$ \SI{1.8}{\kilo\metre}~asl. 
  The majority of the produced fluorescence photons will have a wavelength $\lambda\lesssim$ \SI{400}{\nano\metre}. 
For a zenith angle of \SI{80}{\degree} only $\approx 4~\%$ of the emitted light with wavelength $\lambda =$ \SI{337}{\nano\metre} will reach the observation level, while for \SI{70}{\degree} the observable light increases to $\approx 27\%$ of the emitted light.
  
\subsection{Cherenkov light}

The total number of Cherenkov photons $N^{Ch}$ produced between wavelength $\lambda_1$ and $\lambda_2$ by electrons that are located in a voxel at position $(t, r, \phi)$ is given by
\begin{equation}\label{eq:cherenkov_yield}
    N^{Ch}(t, r, \phi) = \frac{\Delta X(t)}{N_\phi}
    \frac{2\pi}{\alpha\rho(t)} \frac{\lambda_{2} - \lambda_{1}}{\lambda_{1}\lambda_{2}} 
                             \int\limits_{E_{Ch}}^{\epsilon_{\mathrm{max}}}
                             \left(1 - \frac{1}{n^{2}(t) \beta(\epsilon)^{2}}\right) N(t, \epsilon, r/r_M)  ~\mathrm{d}\epsilon
\end{equation}

\noindent where $E_{Ch}$ is the Cherenkov energy-threshold, $\alpha$ the fine-structure constant, $\beta = v/c$,  and $n(t)$ the refractive index\footnote{We replace
the height dependence of refractive index $n$ and specific density $\rho$ with its dependence on $t$}. The produced Cherenkov photons are emitted at the midpoint of the corresponding voxel positions. 

At each voxel and energy bin, the mean angle to the shower axis $\langle \theta_p \rangle$ is calculated from $n_\Omega$ given by \cite{LAFEBRE2009243}.

\begin{figure}[htp!]
\centering
\includegraphics[]{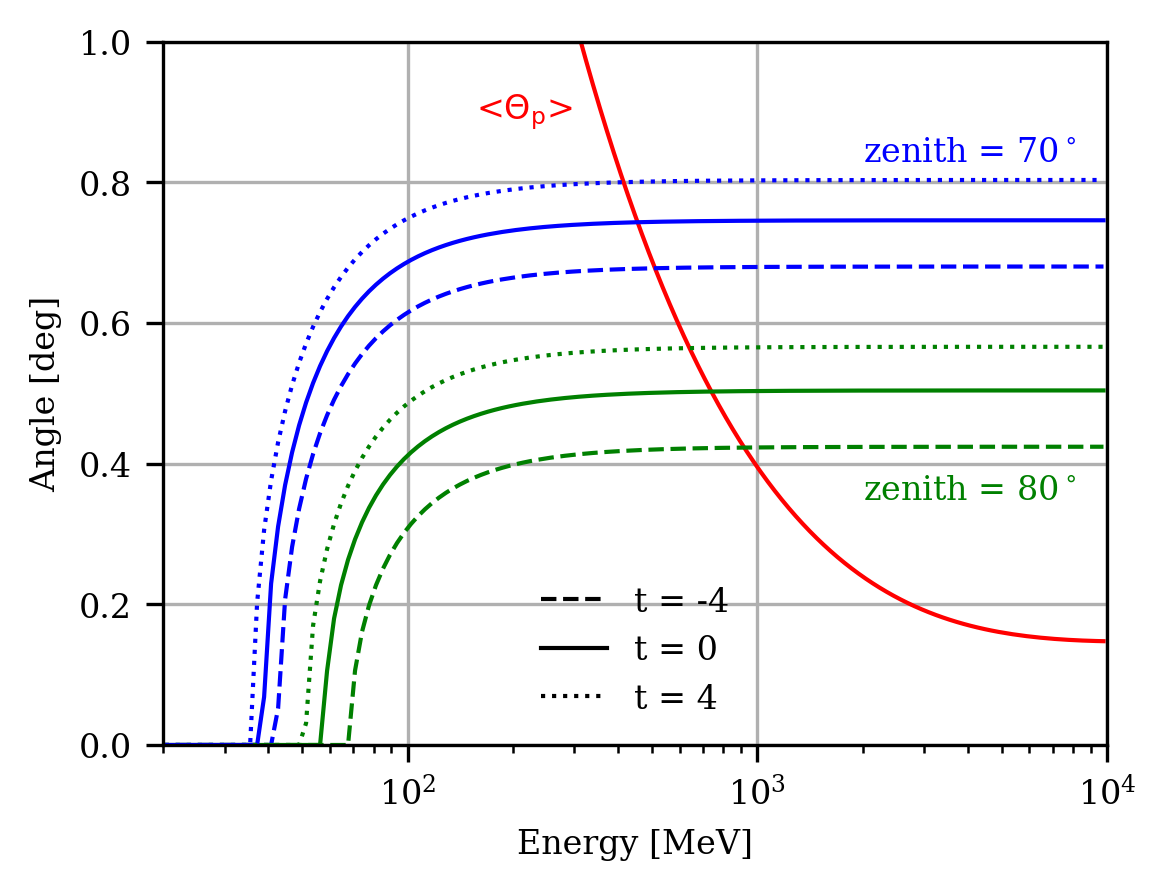}
\caption{The red line marks the mean angle to the shower axis $\langle \theta_{p}\rangle$ 
of electrons as a function of energy. 
Blue and green lines indicate the Cherenkov emission angle for
zenith angles of \SI{70}{\degree} and \SI{80}{\degree} respectively and for
different stages of shower evolution ($t = -4, 0, 4$) marked by  dashed, solid, 
and dotted lines.}
\label{fig:momentum_cherenkov_angle}
\end{figure}

Commonly, the Cherenkov emission angle $\theta_{Ch}$ is assumed to be smaller than 
$\langle \theta_p\rangle$. 
However, for sufficiently large energies, the Cherenkov angle dominates and cannot be neglected anymore.
This is shown in Fig.~\ref{fig:momentum_cherenkov_angle} where
the two angles are compared to each other. The average
angle $\langle \theta_p\rangle$ reaches values similar to the Cherenkov angle at 
energies of $\epsilon=$ \SI{400}{\mega\electronvolt}~$\ldots$~\SI{500}{\mega\electronvolt}. Given that most electrons 
will emit Cherenkov light close to the threshold, it appears at first glance to be a good approximation to neglect the Cherenkov angle. However, even at energies close to 
threshold, the additional widening of the angular distribution of the emitted 
light through the Cherenkov angle 
leads to a substantial change of the order of a few \SI{100}{\metre} on the impact position at the detector level. 

\begin{figure}[htp!]
\includegraphics[trim={1cm 2.3cm 1cm 2cm},clip,width=\linewidth]{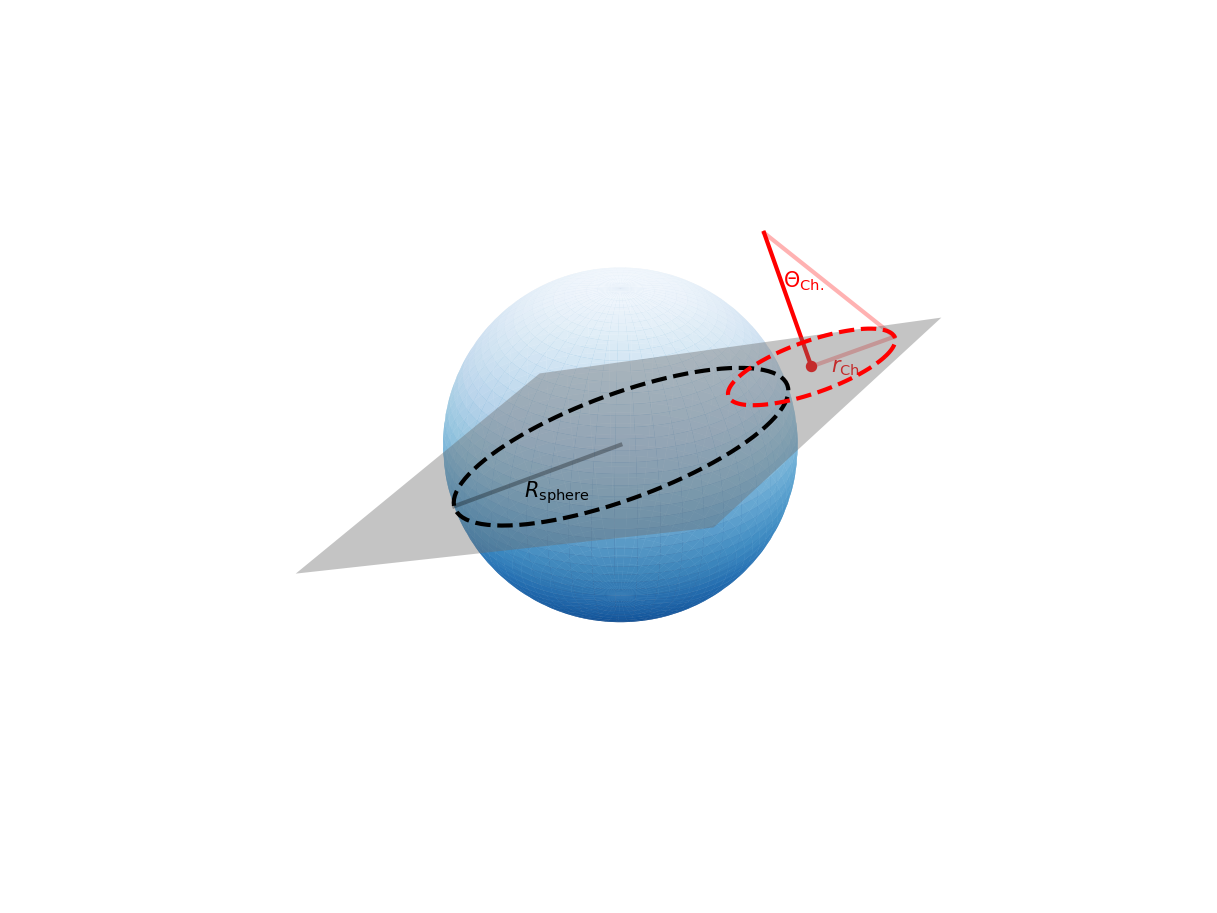}
\caption{Thick red line is parallel to the  momentum vector of the electrons, $\theta_{\mathrm{Ch}}$ the Cherenkov angle and $r_\mathrm{Ch}$ is the radius of the Cherenkov circle in the intersecting plane (in grey).}
\label{fig:cherenkov_detection}
\end{figure}

Finally, the number of Cherenkov photons hitting the telescope's sphere
$N_{sphere}^{Ch}$ is calculated in the following way: 
For each voxel position and energy bin, the equation \ref{eq:cherenkov_yield} is evaluated.
The  momentum vector of the emitting electrons is used to construct a plane 
which is perpendicular
to its direction and intersects the midpoint of the
telescope's sphere. This plane is consequently intersected by the Cherenkov cone and
the telescope's sphere which define two circles (see Fig.~\ref{fig:cherenkov_detection}).
The possible two intersection points mark an arc of length $L$ along the Cherenkov cone
that is inside of the sphere. 
The fraction of  photons intersecting the sphere's surface equals the fraction of the arc length: 

\begin{equation}
    \frac{N_{sphere}^{Ch}}{N^{Ch}} = \frac{L}{2\pi r_{Ch}}.
\end{equation}
This approach does not require any approximations and reduces the necessary computation 
time in comparison to e.g., ray-tracing approaches. 

Note, that in our model particles always move away from the shower axis. For low energy electrons ($\epsilon \sim$ \SI{}{\mega\electronvolt}) this approximation breaks down as shown in Ref.~\cite{LAFEBRE2009243}. However, due to the increased Cherenkov energy threshold $E_{Ch}$ at LZA, these low energy electrons do not contribute to the Cherenkov photon production and can therefore be neglected. Contrary, high energy electrons ($\epsilon \gtrsim$ \SI{50}{\mega\electronvolt}) are produced closer to the shower axis where the majority of the particles are moving away from the shower axis.

\subsection{Imaging of simulated air showers}
\label{sec:imaging}
So far, we have described how we calculate the total number of 
fluorescence and Cherenkov photons produced for each voxel position and how many 
of these photons ($N_{sphere}^{fl}$, $N_{sphere}^{Ch}$)
will intersect the telescope sphere. In this approach, the directional vector of individual photons is not stored, such that the generation of a camera image
requires a simplified treatment. 
To this end,
we make use of the fact that the shower maximum is at a distance $d$ of 
\SIrange{50}{100}{\kilo\metre} which is much larger than the radius of the sphere (e.g., for the largest Cherenkov telescope $R=\SI{14}{\metre}$). We can therefore 
safely assume that all photons from a distant voxel intersect the sphere 
along its diameter. This approximation is justified for large distance $d$, such that the maximum angle difference $\Delta \alpha$ for a photon which is tangential
to the sphere can be approximated  
\begin{equation}
\Delta \alpha = \arctan\left(\frac{R}{d}\right)\approx R/d.
\end{equation}
For the large zenith angles considered here, $d\gtrsim \SI{50}{\kilo\metre}$,
such that $\Delta \alpha \lesssim \SI{0.016}{\degree}$. The resulting widening
of the angular distribution of photons intersecting the sphere is smaller
than the typical angular field of view of individual pixels and can therefore 
be neglected (the smallest pixel  currently used covers a
patch in the sky with a diameter of \SI{0.07}{\degree} \cite{Bolmont_2014}).
This way, the resulting computation time  can be reduced by more than three orders of magnitude in comparison to a ray-tracing approach.
An obvious downside of this method is that the approximation leads
to a systematic underestimate of the image width for the shower tail which even
at large zenith angles is observed at a distance of \SIrange{10}{20}{\kilo\metre}.
For a combination of a very large telescope with a very high-granular camera, the
paraxial approximation introduced breaks down.
Conversely, at smaller zenith angles, the images generated with the paraxial approximation will be only accurate for small-sized telescopes. 
This caveat requires  an individual consideration for each combination
of telescope radius, zenith angle, and pixel size. 

As a final step towards generating realistic camera images which are based on the
number of photo-electrons detected in the individual pixels, we implement
a coarse simulation of the detection process which includes: 
\begin{itemize}
    \item the combined wavelength-dependent quantum efficiency  of the photo-cathode and collection efficiency for photo-electrons,
    \item the wavelength-dependent mirror reflectivity of the mirror facets, and
    \item the  wavelength-dependent transmittance  of the camera protective cover.
\end{itemize}
Additionally the user can provide a single value for all wavelengths which takes into account shadowing effects of the camera support structure and the light guide efficiency of the PMT. 

Therefore, in order to take into account the 
aforementioned wavelength-dependent effects 
in addition to  
atmospheric transmission, 
we pre-compute for each voxel with index $j$ at a given height $h(j)$ 
above ground the efficiency $P_j$ for a photon  
(fluorescence or Cherenkov) to be detected after propagating through the 
atmosphere. The efficiency is calculated by integrating the 
product of emissivity $\Lambda^{Ch,fl}(\lambda)$, detection efficiency $\eta(\lambda)$, and transmission $T(\lambda, h)$. 
The result is normalized
and we obtain: 
\begin{equation}
   P_j =  \frac{\int\limits_{\lambda_{1}}^{\lambda_{2}} \Lambda^{Ch, fl} \cdot T(\lambda, h) \eta(\lambda)~\mathrm{d}\lambda}
   {\int\limits_{\lambda_{1}}^{\lambda_{2}} \Lambda^{Ch, fl}~\mathrm{d}\lambda}.
\end{equation}
The wavelength distribution for the fluorescence photons is given by the wavelength dependent intensity relative to the \SI{337}{\nano\metre} band (see Fig.~\ref{fig:rel_intensity_vs_trans}) and Cherenkov photons are following a $1/\lambda^{2}$ distribution.

The point spread function of the camera is characterized by 
external lookup-tables with the 68\% containment radius as a function of off-axis angle and zenith angle. The night sky background (NSB) can be simulated by providing an average background
rate in units of photo-electrons per nanosecond and per pixel. The detector simulation does not take into account more specific effects related to electronics, for example
formation of a trigger, non-linearities and digitization effects in the read-out. This can be in principle included in this framework or other existing packages like
\texttt{simtel\_array} \cite{Bernl_hr_2008} can be used to obtain a more specific
and accurate simulation of the detector response.

\begin{figure}[htbp!]
\centering
\includegraphics[width=0.7\linewidth]{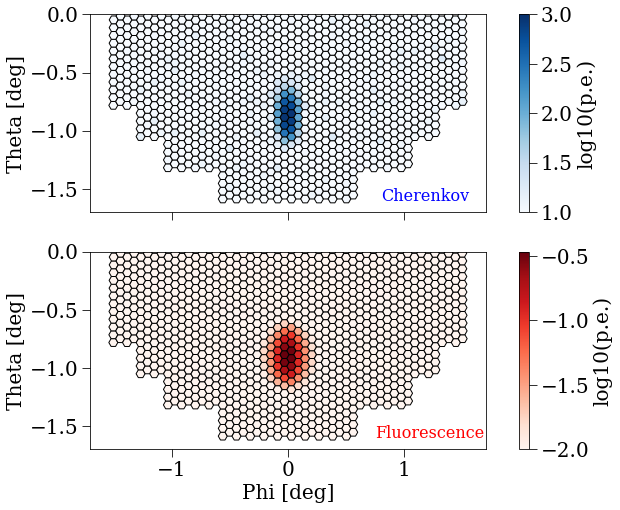}
\caption{Example camera image formed  by Cherenkov  (upper panel) and fluorescence light (lower panel) generated by  an air shower with $E_{\gamma}$ = 780~TeV at a zenith angle of \SI{80}{\degree}. Note, the logarithmic scales differ between
the two images.}
\label{fig:CameraImg}
\end{figure}

The resulting image is shown in Fig.~\ref{fig:CameraImg} formed by Cherenkov (upper panel) and fluorescence light (lower panel) emitted by  an 
gamma-ray air-shower  with an energy 
of $E_\gamma=\SI{780}{\tera\electronvolt}$ at a zenith angle distance of \SI{80}{\degree}. As expected and well-known from observations of air Cherenkov light 
with Cherenkov telescopes, the resulting image resembles an ellipse with 
a major axis (so-called \textit{length}) which is roughly three times as large
as the minor axis (so-called \textit{width}). The fluorescence light image looks very similar to the Cherenkov light image, albeit with a peak intensity reduced by a factor of $\approx 3000$.
Furthermore, the fluorescence light image is of similar length, but noticeably 
larger width than the Cherenkov light image. This is mainly a consequence of the 
fact that Cherenkov emission does have an energy threshold $E_{Ch}$ and that the majority of the particles with $\epsilon > E_{Ch}$ are very close to the shower axis (see Fig.~\ref{fig:energy_dependent_dist}). Contrary, fluorescence emission scales with deposited energy and also electrons which are further away from the shower axis with $\epsilon < E_{Ch}$ will therefore produce fluorescence photons. 
Additionally, fluorescence emission is isotropic while Cherenkov emission does have an opening angle $\theta_{\mathrm{Ch}}$ which limits the visibility of the shower for an IACT.  

\begin{figure}[htbp!]
\centering
\includegraphics[width=0.7\linewidth]{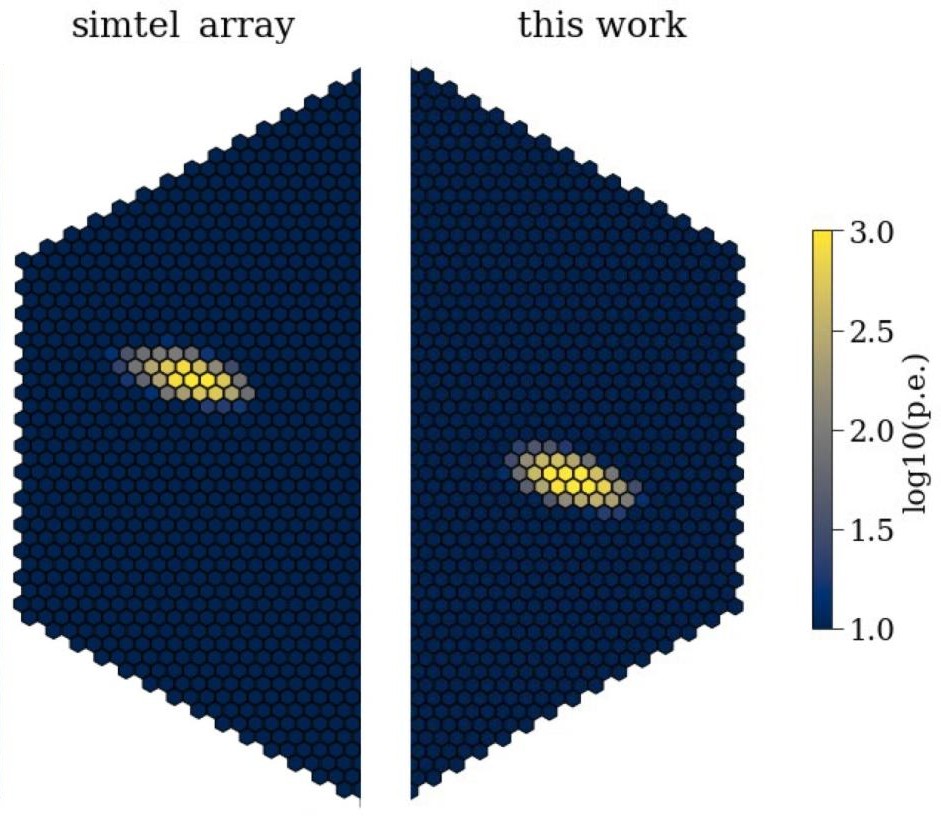}
\caption{Side-by-side comparison of the camera image for a shower with zenith angle of 80$^\circ$ obtained with \texttt{simtel\_array} (left) and \texttt{EASpy} (right). The \texttt{simtel\_array} camera image is rotated by \SI{180}{\degree}.}
\label{fig:Simtel_EASpy_CameraImg}
\end{figure}

\section{Verification with a full simulation}
\label{sec:verification_with_full_simulation}
Finally, we present an example for a shower image in Fig.~\ref{fig:Simtel_EASpy_CameraImg} and compare it side-by-side   
with the shower image as calculated with a complete air-shower simulation 
(see Section~\ref{subsec:corsika}) and ray-tracing for Cherenkov-light using the \texttt{sim\_telarray} package (see Section~\ref{subsec:simtel}). In order to make
the comparison meaningful, we have included the limited dynamical range and saturation of pixel amplitudes similar to \texttt{sim\_telarray}.

Note, that the underlying shower profile is identical for both images. The side-by-side
comparison demonstrates that even with the simplifications introduced above and 
the coarse detector simulation used here, the images appear quite similar. The position of the image in the camera, the light recorded in the pixels are in 
good agreement with each other. The approximate length matches between the two images, while the width appears slightly larger for \texttt{EASpy}. This is readily
explained by the assumption of a symmetric PSF of the 
telescope optics. However, the PSF is in general elongated along the radial direction which leads to a smaller apparent width of the registered image.

To make the comparison more quantitative, we consider a larger sample of
simulated air showers presented in  Fig.~\ref{fig:simulated_showers_energy_dist} and
compare the resulting distributions of width(length)
of the images for on-axis observations for zenith angles of 70° and 80°. 

\subsection{Width distribution}
\label{subsec:width_comparison}
\begin{figure}[htbp!]
     \centering
     \begin{subfigure}[b]{0.47\textwidth}
     \centering
         \includegraphics[width=\textwidth]{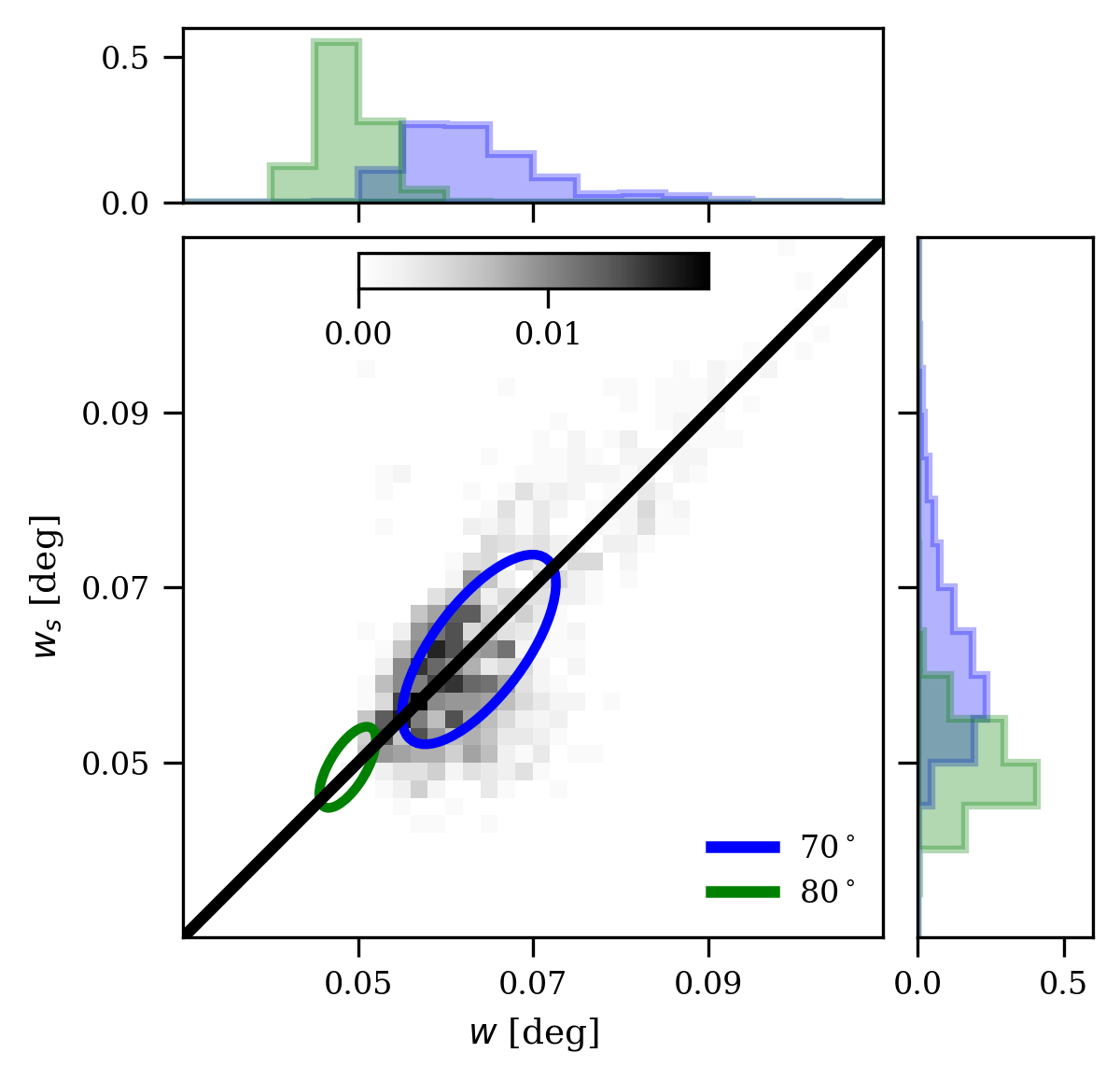}
         \caption{Comparison of width parameter for showers with zenith angle \SI{70}{\degree}(\SI{80}{\degree}) between \texttt{sim\_telarray} ($w_s$ along 
         $y$-axis) and \texttt{EASpy} ($w$ along $x$-axis). The 2D histogram is plotted only for \SI{70}{\degree} for visibility. The blue(green) ellipse describes the 68\% enclosure around the mean value. Top and right panel shows the projection of the $w$ and $w_s$-axis respectively. The black line marks the identity $w=w_s$.}
        \label{fig:width_comparison}
     \end{subfigure}
     \hfill
     \begin{subfigure}[b]{0.47\textwidth}
      \centering
        \includegraphics[width=\textwidth]{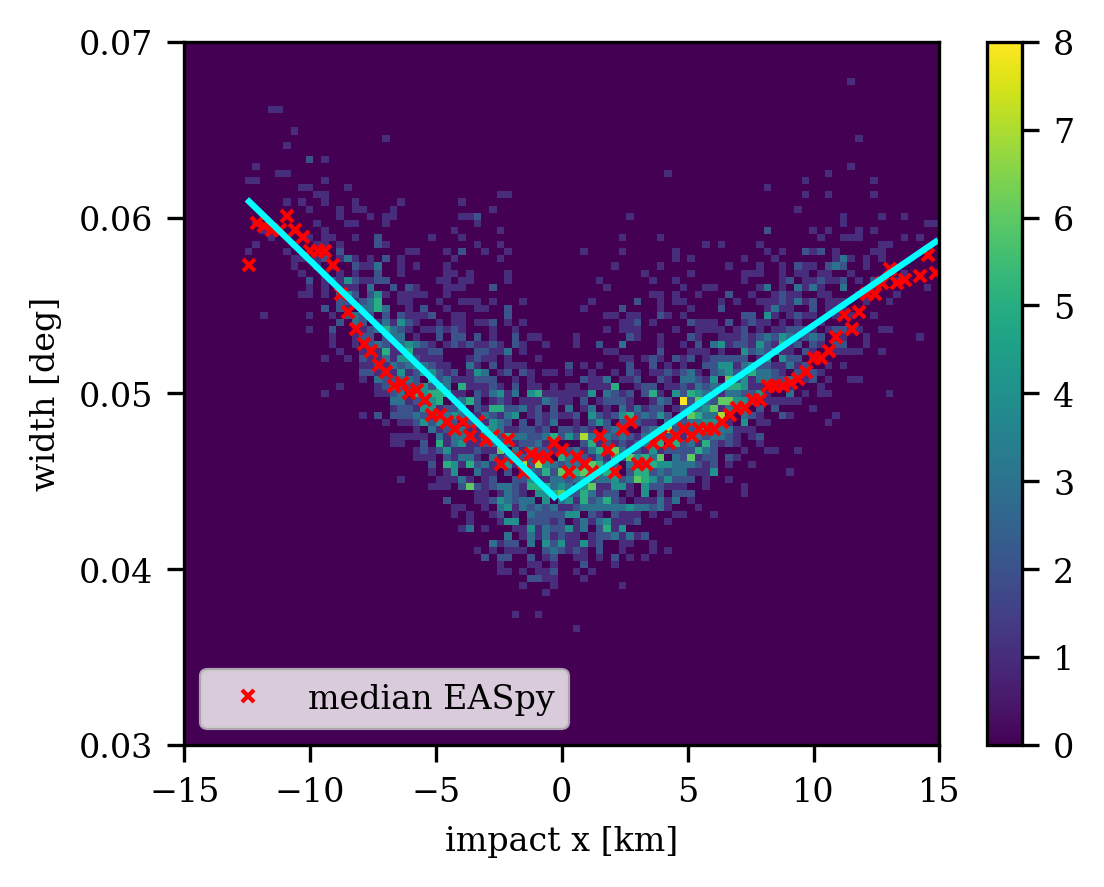}
        \caption{Comparison of width parameter as a function of impact $x$-position for showers with zenith angle \SI{80}{\degree} between \texttt{sim\_telarray} and \texttt{EASpy}. The 2D histogram shows only the values generated with \texttt{sim\_telarray}, the red crosses mark the median values 
        obtained with the \texttt{EASpy}, and the solid lines are derived by fitting a linear function to the median values of \texttt{sim\_telarray} for $x \gtrsim 0$ and $x \lesssim 0$ respectively.}
      \label{fig:TurTur_width}
     \end{subfigure}
\caption{Comparison of width parameter between \texttt{EASpy} and \texttt{sim\_telarray}.}
\end{figure}

In Fig.~\ref{fig:width_comparison}, the image width obtained with the full
simulation ($w_s$) is compared with the results of \texttt{EASpy} ($w$) 
for a zenith angle of \SI{70}{\degree} and \SI{80}{\degree}. 
The two-dimensional histogram for a zenith angle of \SI{70}{\degree} 
indicates a clear and linear correlation between $w_{s}$ and
$w$. The average values as well as the individual distributions are very similar. 
The blue  (\SI{70}{\degree}) and green (\SI{80}{\degree}) ellipses  mark the 68~\% ile boundary around the centroid of the distribution. For each of the two zenith angles, the orientation and position
of the ellipses match well between the two methods. The noticeable  
differences are most likely related to differences in the cleaning procedures that have not been applied in an identical way for the two methods. 
As expected, the width observed at \SI{80}{\degree} is systematically smaller than 
observed at \SI{70}{\degree}, because of the increased distance to the shower maximum.

The excellent agreement between the width $w$ predicted with the 
simplified approach in \texttt{EASpy} and the full simulation encourages further
comparison between image parameters seen at various distances of the telescope
from the shower impact position. In Fig.~\ref{fig:TurTur_width}, we
compare the width values for shower impact positions 
at zenith angle of \SI{80}{\degree} along the $x$-coordinate in the 
observation plane. The telescope is located at $x=y=0$ and the shower moves from 
right to left. The 2D histogram shows the distribution of $w_s$ while the red crosses
are the median values of the $w$-distribution for slices in $x$. For ease of comparison, the cyan line from a linear fit to the $w_s$ distribution is overlaid.  

One can clearly observe consistently for both simulations, 
that with increasing distance $d_{\mathrm{impact}}$ of the telescope position to the impact position of the shower, the width value gets \textit{larger}. At first glance, this appears counter-intuitive since with increasing $d_{\mathrm{impact}}$ the shower should appear \textit{smaller}. However, from a geometrical point of view, the telescope will only register photons with a larger angle with respect to the shower axis for increasing distance $ d_{\mathrm{impact}}$. 
Consequently, the light observed at these distances is emitted from electrons with large $\langle \theta_{p}\rangle$ which on average also have a larger distance to the shower axis. 

The absolute value of the slope of the solid lines in Fig~\ref{fig:TurTur_width} seems to be smaller for $x \gtrsim$ 0 compared to $x \lesssim$ 0. This can be explained by considering that for $x \gtrsim$ 0 the shower maximum is further away from the telescope position compared to $x \lesssim$ 0 and therefore the shower is observed under smaller angles. 

Given these explanations, it is remarkable that the parametrization of the shower 
angle and its dependence on the radial distance to the shower axis is sufficiently 
accurate to capture the interesting property of increasing width with increasing 
distance to the observer. 

\subsection{Length distribution}
\label{subsec:length_comparison}
\begin{figure}[htbp!]
     \centering
     \begin{subfigure}[b]{0.47\textwidth}
     \centering
         \includegraphics[width=\textwidth]{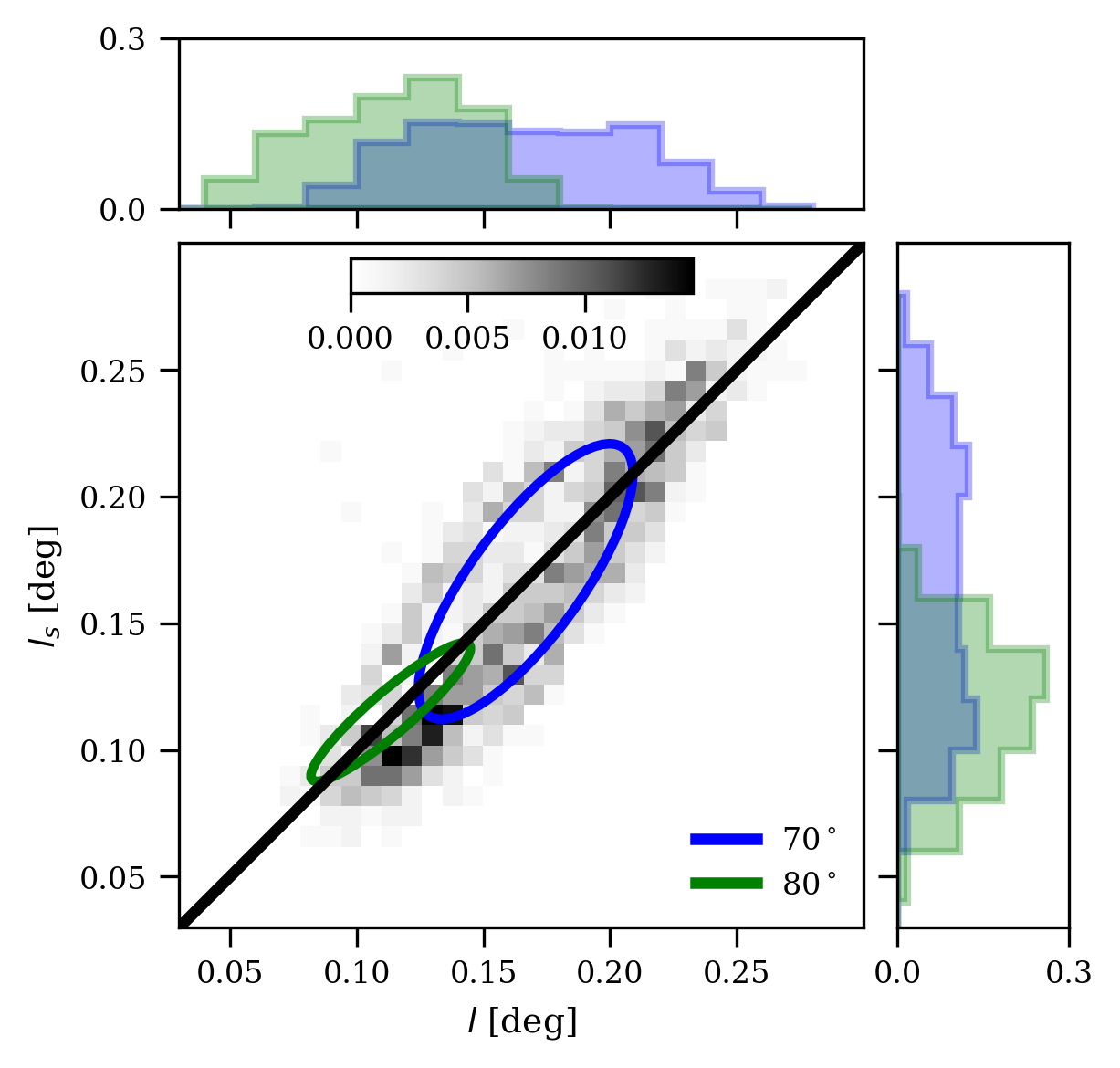}
         \caption{Comparison of length parameter for showers with zenith angle \SI{70}{\degree}(\SI{80}{\degree}) between \texttt{sim\_telarray} ($l$ along $y$-axis) and \texttt{EASpy} ($l_s$ along $x$-axis). The 2D histogram is plotted only for \SI{70}{\degree} for visibility. The blue(green) ellipse describes the 68\% enclosure around the mean value. Top and right panel show the projections of the $l$ and $l_s$-axes respectively. The black line marks the identity $l=l_s$.}
        \label{fig:length_comparison}
     \end{subfigure}
     \hfill
     \begin{subfigure}[b]{0.47\textwidth}
      \centering
      \includegraphics[width=\textwidth]{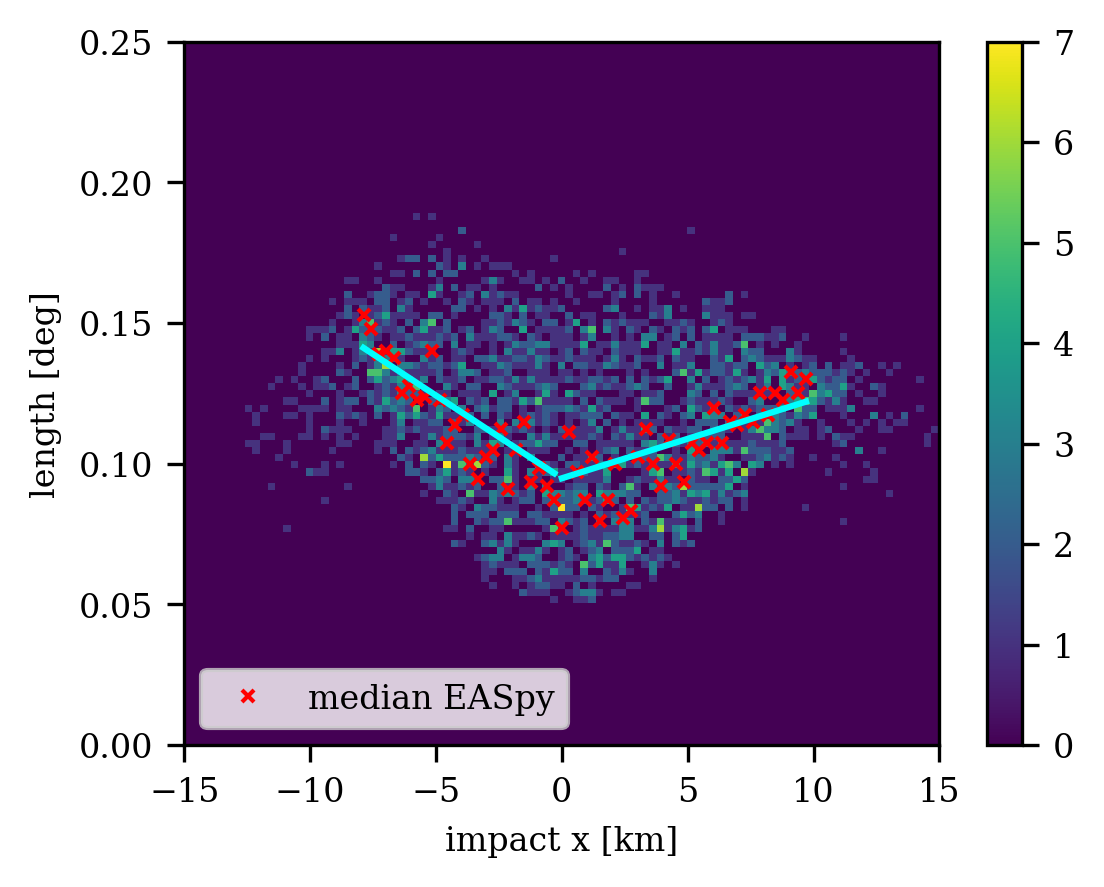}
      \caption{Comparison of length parameter as a function of impact $x$-position for showers with zenith angle \SI{80}{\degree} between \texttt{sim\_telarray} and \texttt{EASpy}. The 2D histogram shows only the values generated with \texttt{sim\_telarray}, the red crosses mark the median values obtained with the \texttt{EASpy}, and the solid lines are derived by fitting a linear function to the median values of \texttt{sim\_telarray} for $x \gtrsim 0$ and $x \lesssim 0$ respectively.}
      \label{fig:TurTur_length}
     \end{subfigure}
\caption{Comparison of length parameter between \texttt{EASpy} and \texttt{sim\_telarray}.}
\end{figure}

The distribution of image length is shown in Fig.~\ref{fig:length_comparison} for
the two considered zenith angles of \SI{70}{\degree} in blue and \SI{80}{\degree} in green.
Overall, the distributions for both zenith angles are in good agreement considering the simplified approach.
 The modest tilt in the orientation of the blue ellipse is caused by a tail of images
 reconstructed with  $l_s\gtrsim \SI{0.18}{\degree}$ for the \texttt{sim\_telarray} distribution. The apparent tilt of the green ellipse highlights the larger values of $l_s$ found for
$l\lesssim \SI{0.11}{\degree}$. 
 This slight deviation is related to the differences in the image cleaning. 
 Within \texttt{sim\_telarray}, pixels are removed with
 amplitudes below a threshold set relative to the brightest pixel whereas in EASpy no
 imaging cleaning is applied (see for a comparison of uncleaned images Appendix~\ref{subsec:simtel}). 
Similar to the width distribution, the average value of the length becomes smaller for
larger zenith angle distances. 

Similar to the presentation of the width parameter in the previous subsection, we compare the apparent length of the images recorded at different positions of
the shower core in the observer plane. The result is shown  in Fig.~\ref{fig:TurTur_length} for a zenith angle of \SI{80}{\degree}.
As before,  the red crosses mark the median values obtained with \texttt{EASpy}
and the overlaid solid lines characterize the  median values of \texttt{sim\_telarray}. 

The conclusions are quite similar to Fig.~\ref{fig:TurTur_width}: for large $ d_{\mathrm{impact}}$ the telescope will only register photons with a large angle with respect to the shower axis leading to an increase of the resulting length parameter. This picture holds as long as the shower is fully captured inside the camera field of view. Once the shower images are getting cut off at the camera edges the length values start to drop (see in Fig.~\ref{fig:TurTur_length} for $x \lesssim$ \SI{-9}{\kilo\metre} and $x \gtrsim$ \SI{10}{\kilo\metre}). The 
close agreement between $l$ and $l_s$ demonstrates the validity of the simplified approach regarding the longitudinal development of the air shower. 

A key difference between Fig.~\ref{fig:TurTur_width} and Fig.~\ref{fig:TurTur_length} is the more uniform spread of values found for 
any given $x$: While for Fig.~\ref{fig:TurTur_width} the difference between the maximum and minimum width value for $x\approx 0$ is $\sim$ \SI{0.01}{\degree}, the corresponding difference for the length value in Fig.~\ref{fig:TurTur_length} is roughly 10 times higher ($\sim$ \SI{0.1}{\degree}). Upon closer inspection, the spread relates to a systematic increase of the length with increasing distance along the $y$-axis for a fixed value of $x$. The obvious explanation is the longitudinal
development of the shower captured in the length parameter while the width parameter is mainly sensitive to the shower maximum.

\section{Photon ground distribution}
\label{sec:ground_distribution}
So far, we have considered the Cherenkov-light images recorded by a telescope observing the
shower with its optical axis aligned parallel to the shower axis. In addition to the Cherenkov light, fluorescence is emitted isotropically and can be observed under arbitrary angles
with respect to the shower axis such that the apparent brightness scales with the inverse
of the distance squared to the shower maximum. At small zenith angles, the maximum photon density of fluorescence and Cherenkov light coincide.

\begin{figure}[htbp!]
\centering
\includegraphics[width=\textwidth]{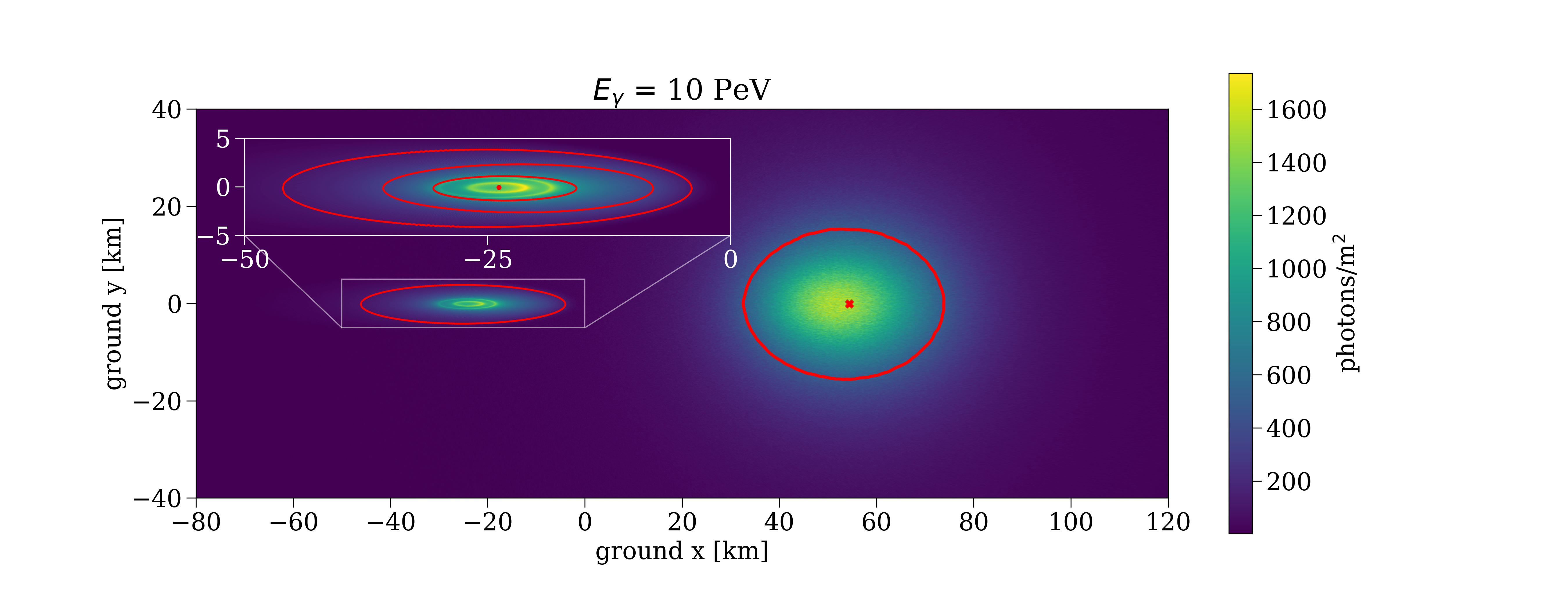}
\caption{Cherenkov and fluorescence photon ground distribution for an air shower with zenith angle \SI{80}{\degree}. The fluorescence photon density is up-scaled by a factor of 5 to match the Cherenkov photon density. The red lines mark an iso-contour at a level of 100~photons/\SI{}{\square\metre} (here the fluorescence density is not up-scaled). Red cross marks the position of the shower maximum. Inset axis: Zoomed-in version of Cherenkov photon ground distribution. The dot marks the position of the impact, red lines indicate the iso-contours at a level of 100~photons/\SI{}{\square\metre}, 350~photons/\SI{}{\square\metre} and 700~photons/\SI{}{\square\metre}.} 
\label{fig:ground_dist}
\end{figure}

 At larger zenith angles, these positions
are displaced with respect to each other. This can be seen in Fig.~\ref{fig:ground_dist}
where both, the photon density of fluorescence and Cherenkov light on the ground (ground level is at a height of \SI{1.8}{\kilo\metre}) are shown
on a grid covering an area of \SI{16000}{\square\kilo\metre} for 
an air shower with $E_\gamma=\SI{10}{\peta\electronvolt}$ at \SI{80}{\degree}. The atmospheric absorption for Cherenkov(fluorescence) photons is treated in a simplified way: we calculate the average absorption from the position of the shower maximum to the ground level for a photon path with zenith angle \SI{80}{\degree}(\SI{0}{\degree}) and apply this factor to all Cherenkov(fluorescence) photons.
The two peak positions are 
displaced by $\approx \SI{80}{\kilo\metre}$. The peak position of the air fluorescence light
pool is close to the projected position of the shower maximum and the peak position of the
Cherenkov light pool is close to but not identical with the shower core position on the ground.

The red ellipses indicate iso-contours of photon surface density of 100~photons/\SI{}{\square\metre}. In this context, it is important to note, that at large zenith angles, 
there is no noticeable contamination of Cherenkov light by fluorescence light. 
This effect has been studied in \cite{Morcuende_2019} for a zenith angle of \SI{30}{\degree}, where the authors show that at distances $\sim$ \SI{1}{\kilo\metre} from the impact 
point 45~\% of the light registered at PeV energies with a wide-angle detector  would be atmospheric fluorescence emission. 

\begin{figure}[htbp!]
\centering
\includegraphics[trim={0.15cm 0cm 0cm 0cm},clip,width=0.7\linewidth]{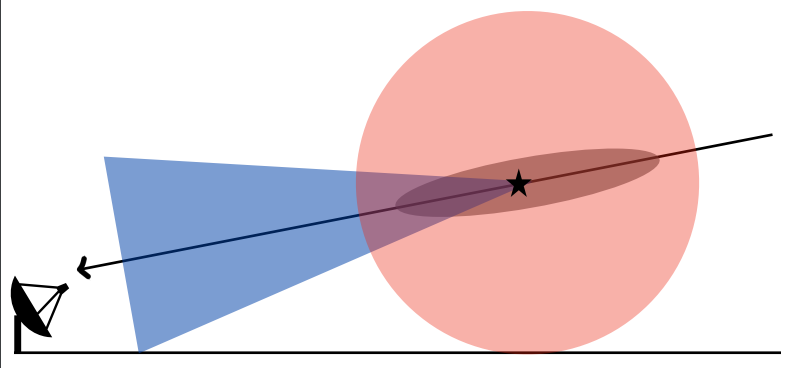}
\caption{Sketch of the geometry properties for Cherenkov and fluorescence emission. The black star marks the shower maximum, the blue shaded area the Cherenkov emission cone and the red shaded area denotes isotropic fluorescence emission. The black line describes the shower axis and the arrow the direction of propagation of the shower.}
\label{fig:fl_ch_sketch}
\end{figure}

While the circular shape of the iso-contour for the fluorescence photon surface density reflects the isotropic nature of fluorescence emission, the iso-contour for the Cherenkov case is elongated in the $x$-direction by  orders of magnitude compared to the $y$-direction. The inset-axis of Fig.~\ref{fig:ground_dist} shows a zoomed-in version of the Cherenkov photon ground distribution, with three isocontours (at 100, 350, and 700~photons/\SI{}{\square\metre}) highlighting its complex structure. Close to the impact position, the
two ring-like structures are directly related to the minimum and maximum
Cherenkov-angle $\theta_{Ch}$ of the superposition of the Cherenkov light emission 
from the entire shower development. Beyond the outer ring feature, the photon density drops 
quickly since this emission is produced mainly be electrons from the later shower development with a large angle with respect
to the shower axes. This drop is more pronounced for light hitting the ground on the far side
of the impact position (in the figure to the left). This is explained by   the geometrical effect of projecting the photon density to the ground (see Fig.~\ref{fig:fl_ch_sketch}). 

\section{Summary}
In this work, we have presented a framework for 
\begin{itemize}
    \item a three-dimensional simulation of extended air showers at large zenith angles,
    \item the consequent emission of fluorescence and Cherenkov light,
    \item and the imaging of the simulated air showers.
\end{itemize}

The three-dimensional simulation of the air shower makes use of parametrizations for electron-positron distributions for particle energy, angular spectrum and lateral distance \cite{LAFEBRE2009243}. Additionally, we assume that these particles are distributed homogeneously around the shower axis such that the simulated air shower follows a cylindrical symmetry.

For the zenith angle range  discussed in this work  (\SIrange[]{70}{80}{\degree}), 
the commonly used approximation of a plane-parallel atmosphere is not applicable anymore. Therefore, we assume a spherical atmosphere in order to accurately calculate the traversed mass density and the corresponding atmospheric parameters along the path of the primary particle. In this framework, we compute  the emission of fluorescence and Cherenkov light in an analytical way on the basis of  M\o ller(Bhabha) cross section and the Frank–Tamm formula
respectively. In this context we note, that at LZA, the Cherenkov emission angle is generally larger than the 
mean angle of electrons/positrons to the shower axis and needs to be included in the calculation.

The amount of light reaching the detector level after Rayleigh- and Mie-scattering as well as absorption processes is determined by using pre-computed lookup-tables generated with the \texttt{MODTRAN} program. A simple plane parallel atmosphere model, where the optical depth scales with the secant of the zenith angle, would underestimate the transmission by $\sim 10$\%. 
Full simulations of Cherenkov and fluorescence photons generated by the particles in the air shower are demanding in terms of computational time and memory required. Here, we 
have introduced a novel approach where the number of photons detected with a telescope are 
determined with geometrical reasoning. As we have demonstrated, this approach is accurate 
for LZA observations and reduces the computational time considerably.  

In order to verify the  approach presented here, we compare 
the width(length) parameter of Cherenkov air-shower images at zenith angle \SI{70}{\degree} and \SI{80}{\degree} between \texttt{EASpy} and \texttt{sim\_telarray}. 
We find for these most relevant imaging parameters very good agreement for the entire
zenith angle range considered.  This is remarkable since the computation time of \texttt{EASpy} does not scale with the energy of the primary particle. In this way, 
\texttt{EASpy} provides a fast and flexible approach to simulate 
a large number of air showers and the
resulting images recorded with various Cherenkov telescope arrays. 
Specifically, for the so-far unexplored energy beyond 100~TeV, it is possible to 
simulate and optimize the sensitivity and performance of these arrays. 

The \texttt{EASpy} framework provides additionally insights into the generation of 
air shower images as well as the distribution of light in the detector plane. We have found that air shower images appear wider and longer with increasing impact distance, an effect that has previously not been noted. This apparent increase is related to the interplay
of the Cherenkov angle and the average angle of shower particles with respect to the shower axis. The effect is confirmed in the detailed simulations using \texttt{CORISKA/sim\_telarray}. 
The distribution of light on the ground calculated with \texttt{EASpy} shows clear separation between the Cherenkov and
fluorescence light pools. This can not be investigated with the standard \texttt{CORSIKA} framework. These features may offer opportunities for observations in future work.

\section*{Acknowledgement}
AB has been funded through the German Ministry for Education and
Research (BMBF) under
contract number 05A20GU3.
We acknowledge the support from the Deutsche Forschungsgemeinschaft (DFG, German Research Foundation) under Germany’s Excellence Strategy – EXC 2121 ``Quantum Universe'' – 390833306.

\appendix
\section{Standard Monte Carlo simulations for imaging Cherenkov telescopes}
\subsection{Air shower simulations with \texttt{CORSIKA}}
\label{subsec:corsika}
The widely used \texttt{CORSIKA} \cite{HeckKnappCapdevielle1998_270043064} 
framework for air shower simulations has been established as the standard software
in the field of ground-based air shower detection. The underlying Monte Carlo approach
has been used in this context since at least 50 years \cite{1974RSPSA.339..133D}. CORSIKA combines the treatment of hadronic and electromagnetic interactions using 
state-of-the art interaction models that have been verified with experimental data
obtained at the highest energy with the large hadron-collider (LHC) \cite{PhysRevC.92.034906}.
The electromagnetic cascade is computed using the well-established EGS-4 code \cite{osti_6137659}. The CORSIKA code includes radiative processes like Cherenkov light generation. Other processes like radio or fluorescence emission are not included in the official release. 

 In \texttt{CORSIKA} charged particles are tracked as they undergo possible interactions or decay, multiple scattering, bending of the trajectory in the Earth`s geomagnetic field, and ionization energy losses, as long as the particle survives a user defined angular and energy cut. 

For the Cherenkov light production each track segment of a charged particle is divided into smaller sub-steps such that in each sub-step one photon \textit{bunch} is emitted at the midpoint of the sub-step. The number of Cherenkov photons in one bunch is given by the user defined \textit{bunchsize} which defines the number of sub-steps needed until the number of Cherenkov photons in one bunch is less than the defined bunchsize.
For our simulation, we set the \textit{bunchsize} to five photons which is a compromise between resolution and computation time required. For the typical 
detection efficiency of $\approx 10~\%$ for ground-based detection, less than
one photon would be detected per bunch. 

At each step, the number of produced Cherenkov photons and the emission angle is determined by the height-dependent refractive index and velocity of the charged particle, while assuming a continuous energy loss along its path. Within the IACT package for \texttt{CORSIKA}, telescopes are modelled as spheres with adjustable position and radius for each telescope. The bunches are transported to ground level and for those intersecting a telescope sphere, 
all relevant informations are stored in a file.
Additionally, \texttt{CORSIKA} provides the option to count the 
number of particles passing through intervals of slant depth and their energy deposit for various particle groups ($\gamma$, $e^{\pm}$, $\mu^{\pm}$, hadrons, Cherenkov photons) This information is written to a separate file. 

In this work we used \texttt{CORSIKA} (v7.7402)
to obtain shower profiles for photon air showers with energies in the range of \SIrange{100}{1000}{\tera\electronvolt } and zenith angles between \SI{70}{\degree} and \SI{80}{\degree} in steps of \SI{2}{\degree}.

\begin{figure}[htbp!]
     \centering
     \begin{subfigure}[b]{0.49\textwidth}
     \centering
     \includegraphics[]{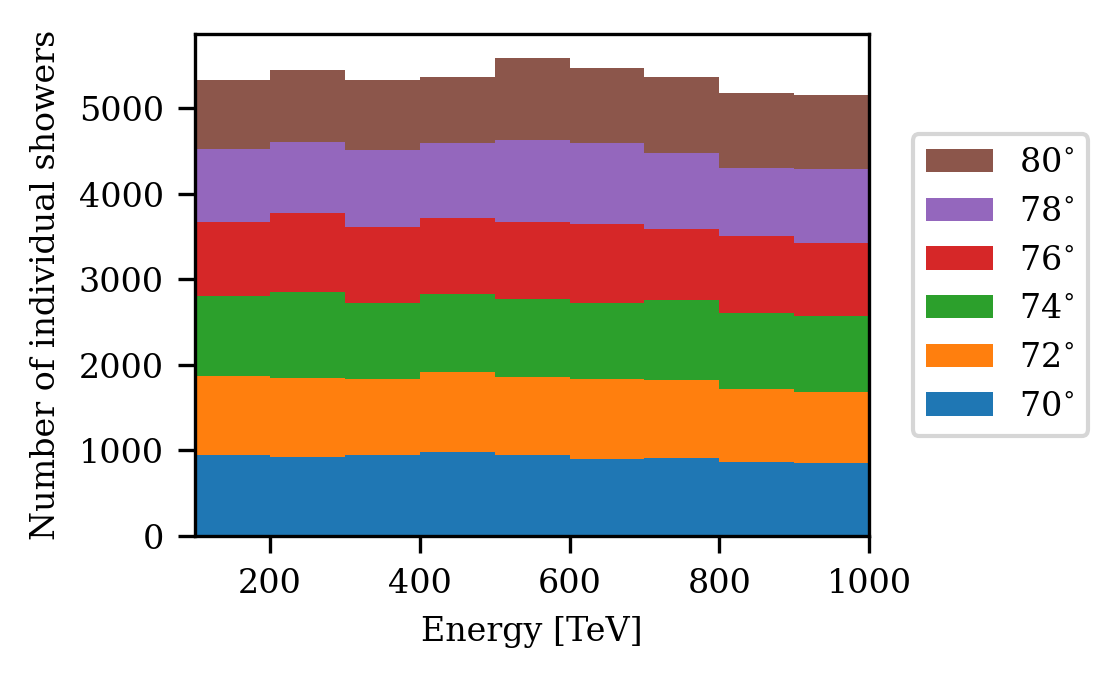}
     \caption{}
     \label{fig:simulated_showers_energy_dist}
     \end{subfigure}
     \hfill
     \begin{subfigure}[b]{0.49\textwidth}
      \centering
      \includegraphics[]{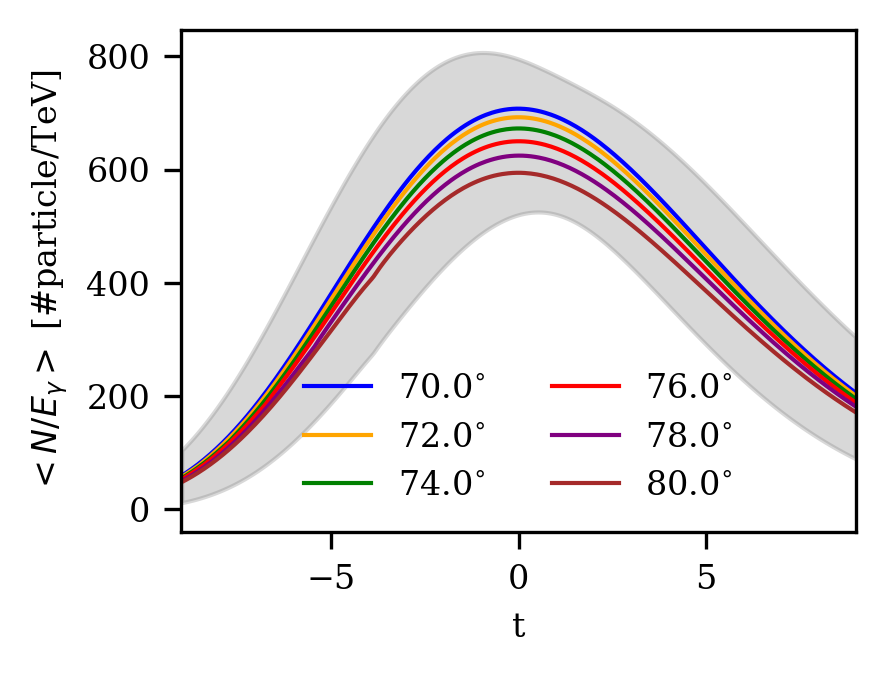}
      \caption{}
      \label{fig:profiles}
     \end{subfigure}
\caption{(a) Histogram of simulated showers with bin width of \SI{100}{\tera\electronvolt} for each zenith angle. (b) Mean number of particles per primary gamma energy for each zenith angle bin. The gray shaded area denotes the envelope of the 1~$\sigma$ error bands of zenith angle bins. }
\end{figure}

The resulting shower profiles are used for calculating the electron-positron distributions for particle energy, angular spectrum and lateral distance (see Section \ref{sec:parametrizing}). For each zenith angle, we simulated $\sim 8\,000$ showers which sums up to a total of $\sim 50\,000$ showers (see Fig. \ref{fig:simulated_showers_energy_dist}). 

The corresponding mean shower profile per energy of the primary photon for each zenith angle bin is shown in Fig. \ref{fig:profiles}. The shower profiles are described in terms of the relative evolution stage $t$ (see Eq. \ref{eq:relative_evolution_stage_t}). While the shower profiles converge at very early ($t \leq$ -4) and late stages ($t \geq$ 4), they surprisingly deviate around the shower maximum ($t \sim 0$). This behavior is not seen, if the shower profiles are normalized to the number of particles at the shower maximum $N/N_{\mathrm{max}}$, with $N_{\mathrm{max}} = N(t = 0)$. 

In the following, we provide all relevant information regarding \texttt{CORSIKA} version, compiling options, and parameter settings used to simulate the air showers:

\begin{description}
   \item[\texttt{CORSIKA} version] 7.7402
   \item[compiling options] VOLUMEDET, IACT, ATMEXT, CURVED, VIEWCONE, SLANT, CERENKOV
   \item[interaction models] QGSJET-01D \cite{KALMYKOV199717}, GHEISHA 2002d \cite{GHEISHA}, EGS4 \cite{osti_6137659}
   \item[atmosphere] all-year average profile at H.E.S.S. site
\end{description}

\noindent For the energy cut-off of the particle kinetic energy for hadrons, muons and electrons we used \SI{0.3}{\giga\electronvolt}, \SI{0.1}{\giga\electronvolt} and \SI{0.001}{\giga\electronvolt} respectively. The lower and upper limit of the wavelength band for the Cherenkov radiation production covers a range \SIrange[]{250}{700}{\nano\metre} with a bunchsize of 5. The longitudinal development of particle numbers and energy deposit by ionization energy losses are sampled in steps of $2~\mathrm{g/cm^{2}}$. Explanations for the compiling options and keywords for the \texttt{CORSIKA} input card can be found in \cite{CORSIKA_users_guide}.

\subsection{Detector simulation with sim\_telarray}
\label{subsec:simtel}
The shower simulation as described in the previous section provides the 
information on the Cherenkov light received by a telescope located in the center
of the so-called telescope sphere. The subsequent treatment of the different components of a telescope require a dedicated simulation. Since this simulation requires in-depth and specific details of the telescope hardware, it has often been 
treated by custom codes that are limited to a single experiment. In the past decade, 
\texttt{sim\_telarray} \cite{Bernl_hr_2008} has been used for the HEGRA, H.E.S.S., and 
 planned CTA IACT systems. 

With the \texttt{sim\_telarray} package each telescope of an IACT system can be configured individually. The photon information stored by CORSIKA are read-in and 
the atmospheric transmission defined by an external look-up table is applied to
the photon bunches.
Using ray-tracing, the remaining Cherenkov light is followed through the various optical components, until it converts into a photo-electron on the photocathode. The optical
components can be configured in a flexible way to take into account the geometry 
of the mirrors, their alignment as well as the overall layout of the camera to include
shadowing effects of masts etc. 
The efficiencies for reflection, absorption, and conversion are treated in a wave-length
dependent way. The pulse-shape, after-pulsing, relative timing, triggering, and digitization can be fine-tuned with various parameters to match the actual setup. 
In addition to the Cherenkov light, the night-sky background is included as a
noise source to the camera images.

Information about the position and number of photo-electrons for every pixel enables geometrical reconstruction of the shower using the classical Hillas parameters technique. While the \texttt{sim\_telarray} package itself does not provide any reconstruction steps which involve look-up of other simulation data, e.g. gamma-hadron separation, it does provide a Postscript-generator for visualisation of the data, which includes the moments of the Hillas ellipse and the reconstructed shower direction.  

\begin{figure}[htbp!]
\centering
\includegraphics[]{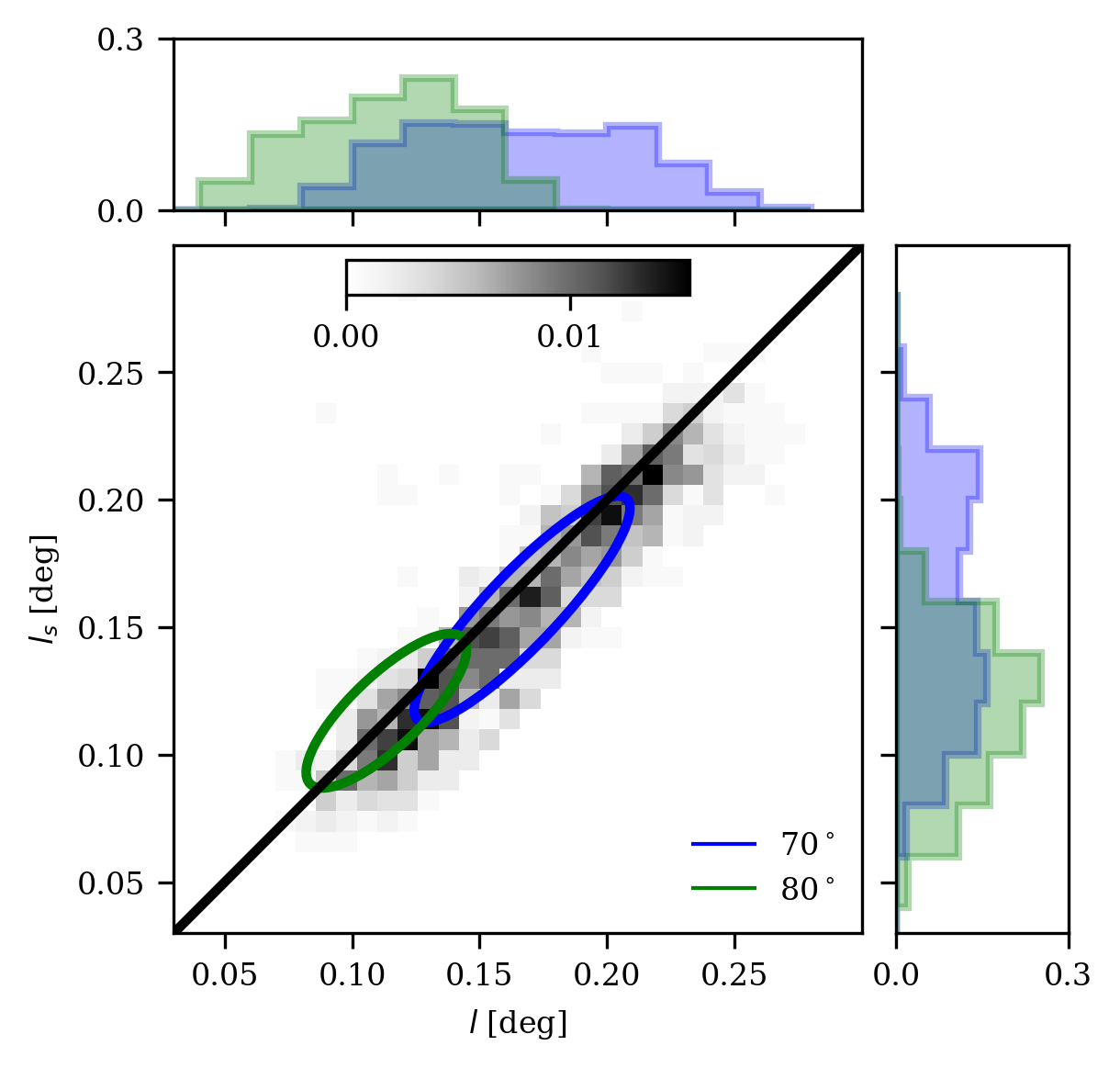}
\caption{Same as in Fig.~\ref{fig:length_comparison} but here using the uncleaned images from \texttt{sim\_telarray}.}
\label{fig:length_comparison_uncleaned}
\end{figure}

As mentioned in section~\ref{subsec:length_comparison}, \texttt{sim\_telarray} applies a cleaning procedure in order to remove photo-electrons from the night sky background and electronics. This procedure will also inevitably remove photo-electrons produced by Cherenkov-photons. The small effects of the cleaning procedure can be observed by comparing Fig.~\ref{fig:length_comparison} with Fig.~\ref{fig:length_comparison_uncleaned}, where in Fig.~\ref{fig:length_comparison_uncleaned} the uncleaned images (i.e. only the photo-electrons produced by Cherenkov-photons) from \texttt{sim\_telarray} were used.

\section{Fluorescence parameters}
\label{sec:appendix_FL}
In Table~\ref{tab:appendix_Fl} we provide all necessary parameter values in order to evaluate $Y_{air}(\lambda, T, p, p_{w})$ in Eq.~\ref{eq:dNdX_fl}. The parameter values are taken from \cite{Ave_2007}, \cite{Ave_2008} and \texttt{ShowerModel}.

\begin{longtable}{|ccccc|}
\caption{Parameters for Fluorescence model. Reference atmospheric conditions for dry air are $p_{0} = 800.0$~hPa and $T_{0} = 293.0$~K. Reference yield of the \SI{337}{\nano\metre} band is $Y_{air}(337, T_{0}, p_{0}) = 7.04$~photons/MeV. The table describes from left to right: wavelength $\lambda$ in units of nm, relative intensity of each wavelength at reference atmospheric conditions, quenching pressure of dry air $PP0$ and quenching pressure of water vapor $PPw$ in units of hPa and a parameter $a$ which takes into account the temperature dependence.}
\label{tab:appendix_Fl}\\
\hline
$\lambda$ & $\frac{I_{\lambda}(T_{0}, p_{0})}{I_{337}(T_{0},p_{0})}$ & $PP0$ & $PPw$ & $a$   \\
\endfirsthead
\endhead
\hline
\endfoot
\endlastfoot
nm        & $\times 10^{-2}$                                         & hPa   & hPa   & -     \\ \hline
296       & 5.16                                                     & 18.50 & 0.00  & 0.00  \\
298       & 2.77                                                     & 17.30 & 0.00  & 0.00  \\
302       & 0.41                                                     & 21.00 & 0.00  & 0.00  \\
308       & 1.44                                                     & 21.00 & 0.00  & 0.00  \\
312       & 7.24                                                     & 18.70 & 0.00  & 0.00  \\
314       & 11.05                                                    & 12.27 & 1.20  & -0.13 \\
316       & 39.33                                                    & 11.88 & 1.10  & -0.19 \\
318       & 0.46                                                     & 21.00 & 0.00  & 0.00  \\
327       & 0.80                                                     & 19.00 & 0.00  & 0.00  \\
329       & 3.80                                                     & 20.70 & 0.00  & 0.00  \\
331       & 2.15                                                     & 16.90 & 0.00  & 0.00  \\
334       & 4.02                                                     & 15.50 & 0.00  & 0.00  \\
337       & 100.0                                                    & 15.89 & 1.28  & -0.35 \\
346       & 1.74                                                     & 21.00 & 0.00  & 0.00  \\
350       & 2.79                                                     & 15.20 & 1.50  & -0.38 \\
354       & 21.35                                                    & 12.70 & 1.27  & -0.22 \\
358       & 67.41                                                    & 15.39 & 1.30  & -0.35 \\
366       & 1.13                                                     & 21.00 & 0.00  & 0.00  \\
367       & 0.54                                                     & 19.00 & 0.00  & 0.00  \\
371       & 4.97                                                     & 14.80 & 1.30  & -0.24 \\
376       & 17.87                                                    & 12.82 & 1.10  & -0.17 \\
381       & 27.20                                                    & 16.51 & 1.40  & -0.34 \\
386       & 0.5                                                      & 19.00 & 0.00  & 0.00  \\
388       & 1.17                                                     & 7.60  & 0.00  & 0.00  \\
389       & 0.83                                                     & 3.90  & 0.00  & 0.00  \\
391       & 28.00                                                    & 2.94  & 0.33  & -0.79 \\
394       & 3.36                                                     & 13.70 & 1.20  & -0.20 \\
400       & 8.38                                                     & 13.60 & 1.10  & -0.20 \\
405       & 8.07                                                     & 17.80 & 1.50  & -0.37 \\
414       & 0.49                                                     & 19.0  & 0.00  & 0.00  \\
420       & 1.75                                                     & 13.80 & 0.00  & 0.00  \\
424       & 1.04                                                     & 3.90  & 0.00  & 0.00  \\
427       & 7.08                                                     & 6.38  & 0.00  & 0.00  \\
428       & 4.94                                                     & 2.89  & 0.60  & -0.54 \\ \hline
\end{longtable}




\bibliographystyle{JHEP}
\bibliography{biblio.bib}

\providecommand{\href}[2]{#2}\begingroup\raggedright\begin{thebibliography}{10}

\bibitem{2023A&A...671A..67D}
L.~{Dirson} and D.~{Horns}, \emph{{Phenomenological modelling of the Crab
  Nebula's broadband energy spectrum and its apparent extension}},
  \href{https://doi.org/10.1051/0004-6361/202243578}{\emph{A\&A} {\bfseries
  671} (2023) A67} [\href{https://arxiv.org/abs/2203.11502}{{\ttfamily
  2203.11502}}].

\bibitem{2011ExA....32..193A}
M.~{Actis}, G.~{Agnetta}, F.~{Aharonian}, A.~{Akhperjanian}, J.~{Aleksi{\'c}},
  E.~{Aliu} et~al., \emph{{Design concepts for the Cherenkov Telescope Array
  CTA: an advanced facility for ground-based high-energy gamma-ray astronomy}},
  \href{https://doi.org/10.1007/s10686-011-9247-0}{\emph{Experimental
  Astronomy} {\bfseries 32} (2011) 193}
  [\href{https://arxiv.org/abs/1008.3703}{{\ttfamily 1008.3703}}].

\bibitem{Sommer_et_al}
P.~{Sommers} and J.W.~{Elbert}, \emph{{Ultra-high-energy gamma-ray astronomy
  using atmospheric Cerenkov detectors at large zenith angles}},
  \href{https://doi.org/10.1088/0305-4616/13/4/019}{\emph{Journal of Physics G
  Nuclear Physics} {\bfseries 13} (1987) 553}.

\bibitem{MAGIC_LZA_CRAB2020}
M.~Collaboration, \emph{{MAGIC} very large zenith angle observations of the
  crab nebula up to 100 tev},
  \href{https://doi.org/10.1051/0004-6361/201936899}{\emph{Astronomy \&
  Astrophysics} {\bfseries 635} (2020) A158}.

\bibitem{DESOUZA2004263}
V.~{de Souza}, H.~Barbosa and C.~Dobrigkeit, \emph{A monte carlo method to
  generate fluorescence light in extensive air showers},
  \href{https://doi.org/https://doi.org/10.1016/j.astropartphys.2004.07.006}{\emph{Astroparticle
  Physics} {\bfseries 22} (2004) 263}.

\bibitem{Morcuende_2019}
D.~Morcuende, J.~Rosado, J.~Contreras and F.~Arqueros, \emph{Relevance of the
  fluorescence radiation in {VHE} gamma-ray observations with the cherenkov
  technique},
  \href{https://doi.org/10.1016/j.astropartphys.2018.11.003}{\emph{Astroparticle
  Physics} {\bfseries 107} (2019) 26}.

\bibitem{Greisen}
K.~{Greisen}, \emph{{Cosmic Ray Showers}},
  \href{https://doi.org/10.1146/annurev.ns.10.120160.000431}{\emph{Annual
  Review of Nuclear and Particle Science} {\bfseries 10} (1960) 63}.

\bibitem{Gaisser_Hillas}
T.K.~{Gaisser} and A.M.~{Hillas}, \emph{{Reliability of the Method of Constant
  Intensity Cuts for Reconstructing the Average Development of Vertical
  Showers}},  in \emph{International Cosmic Ray Conference}, vol.~8 of
  \emph{International Cosmic Ray Conference}, p.~353, Jan., 1977.

\bibitem{Nishimura_Kamata}
K.~{Kamata} and J.~{Nishimura}, \emph{{The Lateral and the Angular Structure
  Functions of Electron Showers}},
  \href{https://doi.org/10.1143/PTPS.6.93}{\emph{Progress of Theoretical
  Physics Supplement} {\bfseries 6} (1958) 93}.

\bibitem{Nerling_2006}
F.~Nerling, J.~Blümer, R.~Engel and M.~Risse, \emph{Universality of electron
  distributions in high-energy air showers{\textemdash}description of cherenkov
  light production},
  \href{https://doi.org/10.1016/j.astropartphys.2005.09.002}{\emph{Astroparticle
  Physics} {\bfseries 24} (2006) 421}.

\bibitem{jaime_rosado_2022_6773258}
J.~Rosado and D.~Morcuende, \emph{Jaimerosado/showermodel: v0.1.9},  June,
  2022.
\newblock 10.5281/zenodo.6773258.

\bibitem{Morcuende:2021bmo}
D.~Morcuende and J.~Rosado, \emph{{ShowerModel: A Python Package for Modelling
  Cosmic-ray Showers, Their Light Production and Their Detection}}, {\emph{ASP
  Conf. Ser.} {\bfseries 532} (2022) 155}
  [\href{https://arxiv.org/abs/2103.00578}{{\ttfamily 2103.00578}}].

\bibitem{MODTRAN}
F.K.~et~al., \emph{The MODTRAN 2/3 Report and LOW-TRAN 7 Model}, Phillips
  Laboratory, Hanscom AFB, MA 01731, USA (1996).

\bibitem{LAFEBRE2009243}
S.~Lafebre, R.~Engel, H.~Falcke, J.~Hörandel, T.~Huege, J.~Kuijpers et~al.,
  \emph{Universality of electron–positron distributions in extensive air
  showers},
  \href{https://doi.org/https://doi.org/10.1016/j.astropartphys.2009.02.002}{\emph{Astroparticle
  Physics} {\bfseries 31} (2009) 243}.

\bibitem{SELTZER19821189}
S.M.~Seltzer and M.J.~Berger, \emph{Evaluation of the collision stopping power
  of elements and compounds for electrons and positrons},
  \href{https://doi.org/https://doi.org/10.1016/0020-708X(82)90244-7}{\emph{The
  International Journal of Applied Radiation and Isotopes} {\bfseries 33}
  (1982) 1189}.

\bibitem{doi:10.1146/annurev.ns.04.120154.001531}
E.A.~Uehling, \emph{Penetration of heavy charged particles in matter},
  \href{https://doi.org/10.1146/annurev.ns.04.120154.001531}{\emph{Annual
  Review of Nuclear Science} {\bfseries 4} (1954) 315}
  [\href{https://arxiv.org/abs/https://doi.org/10.1146/annurev.ns.04.120154.001531}{{\ttfamily
  https://doi.org/10.1146/annurev.ns.04.120154.001531}}].

\bibitem{PDG}
P.D.~Group, \emph{Particle physics booklet},  2022.

\bibitem{CORSIKA_users_guide}
T.P.~D.~Heck, \emph{Extensive air shower simulation with corsika: A user’s
  guide, version 7.7410},  2021.

\bibitem{Arqueros_2009}
F.~Arqueros, F.~Blanco and J.~Rosado, \emph{Analysis of the fluorescence
  emission from atmospheric nitrogen by electron excitation, and its
  application to fluorescence telescopes},
  \href{https://doi.org/10.1088/1367-2630/11/6/065011}{\emph{New Journal of
  Physics} {\bfseries 11} (2009) 065011}.

\bibitem{Itikawa_et_al}
Y.~Itikawa, M.~Hayashi, A.~Ichimura, K.~Onda, K.~Sakimoto, K.~Takayanagi
  et~al., \emph{Cross sections for collisions of electrons and photons with
  nitrogen molecules}, \href{https://doi.org/10.1063/1.555762}{\emph{Journal of
  Physical and Chemical Reference Data} {\bfseries 15} (1986) 985}
  [\href{https://arxiv.org/abs/https://doi.org/10.1063/1.555762}{{\ttfamily
  https://doi.org/10.1063/1.555762}}].

\bibitem{Ave_2007}
M.~Ave, M.~Bohacova, B.~Buonomo, N.~Busca, L.~Cazon, S.~Chemerisov et~al.,
  \emph{Measurement of the pressure dependence of air fluorescence emission
  induced by electrons},
  \href{https://doi.org/10.1016/j.astropartphys.2007.04.006}{\emph{Astroparticle
  Physics} {\bfseries 28} (2007) 41}.

\bibitem{Ave_2008}
M.~Ave, M.~Bohacova, B.~Buonomo, N.~Busca, L.~Cazon, S.~Chemerisov et~al.,
  \emph{Temperature and humidity dependence of air fluorescence yield measured
  by {AIRFLY}}, \href{https://doi.org/10.1016/j.nima.2008.08.050}{\emph{Nuclear
  Instruments and Methods in Physics Research Section A: Accelerators,
  Spectrometers, Detectors and Associated Equipment} {\bfseries 597} (2008)
  50}.

\bibitem{Bolmont_2014}
J.~Bolmont, P.~Corona, P.~Gauron, P.~Ghislain, C.~Goffin, L.G.~Riveros et~al.,
  \emph{The camera of the fifth h.e.s.s. telescope. part i: System
  description}, \href{https://doi.org/10.1016/j.nima.2014.05.093}{\emph{Nuclear
  Instruments and Methods in Physics Research Section A: Accelerators,
  Spectrometers, Detectors and Associated Equipment} {\bfseries 761} (2014)
  46}.

\bibitem{Bernl_hr_2008}
K.~Bernlöhr, \emph{Simulation of imaging atmospheric cherenkov telescopes with
  {CORSIKA} and sim{\_}telarray},
  \href{https://doi.org/10.1016/j.astropartphys.2008.07.009}{\emph{Astroparticle
  Physics} {\bfseries 30} (2008) 149}.

\bibitem{HeckKnappCapdevielle1998_270043064}
D.~Heck, J.~Knapp, J.N.~Capdevielle, G.~Schatz and T.~Thouw, \emph{Corsika: A
  monte carlo code to simulate extensive air showers},  Tech. Rep. (1998),
  \href{https://doi.org/10.5445/IR/270043064}{DOI}.

\bibitem{1974RSPSA.339..133D}
H.E.~{Dixon}, J.C.~{Earnshaw}, J.R.~{Hook}, J.H.~{Hough}, G.J.~{Smith},
  W.~{Stephenson} et~al., \emph{{Computer Simulations of Cosmic-Ray Air
  Showers. I. Average Characteristics of Proton Initiated Showers}},
  \href{https://doi.org/10.1098/rspa.1974.0114}{\emph{Proceedings of the Royal
  Society of London Series A} {\bfseries 339} (1974) 133}.

\bibitem{PhysRevC.92.034906}
T.~Pierog, I.~Karpenko, J.M.~Katzy, E.~Yatsenko and K.~Werner, \emph{Epos lhc:
  Test of collective hadronization with data measured at the cern large hadron
  collider}, \href{https://doi.org/10.1103/PhysRevC.92.034906}{\emph{Phys. Rev.
  C} {\bfseries 92} (2015) 034906}.

\bibitem{osti_6137659}
W.R.~Nelson, H.~Hirayama and D.W.~Rogers, \emph{Egs4 code system}, .

\bibitem{KALMYKOV199717}
N.~Kalmykov, S.~Ostapchenko and A.~Pavlov, \emph{Quark-gluon-string model and
  eas simulation problems at ultra-high energies},
  \href{https://doi.org/https://doi.org/10.1016/S0920-5632(96)00846-8}{\emph{Nuclear
  Physics B - Proceedings Supplements} {\bfseries 52} (1997) 17}.

\bibitem{GHEISHA}
H.~Fesefeldt, \emph{The simulation of hadronic shower - physics and
  applications, pitha-85/02, rwth aachen},  1985.

\end{thebibliography}\endgroup
\end{document}